\documentclass{article}

\usepackage[utf8]{inputenc}
\DeclareUnicodeCharacter{2212}{-}
\usepackage{amsmath}
\usepackage{amssymb}
\usepackage{amsfonts}
\usepackage{pifont}

\usepackage[backend=bibtex,
style=numeric,
bibencoding=ascii,
maxbibnames=99,
]{biblatex}
\usepackage[compact]{titlesec}
\usepackage[T1]{fontenc}    %
\usepackage{enumitem}

\usepackage{booktabs}       %
\usepackage{amsfonts}       %
\usepackage{nicefrac}       %
\usepackage{microtype}      %
\usepackage{xcolor}         %
\usepackage{listings}
\usepackage{dashrule}
\usepackage{wrapfig}

\usepackage{amsmath,amsfonts,bm}

\def\eqref#1{equation~\ref{#1}}

\def\1{\bm{1}}

\DeclareMathAlphabet{\mathsfit}{\encodingdefault}{\sfdefault}{m}{sl}
\SetMathAlphabet{\mathsfit}{bold}{\encodingdefault}{\sfdefault}{bx}{n}

\usepackage{amsfonts}

\usepackage[breaklinks=true,colorlinks,citecolor=black,bookmarks=false]{hyperref}
\hypersetup{
    colorlinks=true,
	linkcolor=blue,
	filecolor=magenta,      
	urlcolor=blue,
	citecolor=black,
	pdfinfo={
		Title={X-Risk Analysis for AI Research},
		Author={Dan Hendrycks, Mantas Mazeika},
		Subject={ML Safety, AI X-Risk},
		Keywords={x-risk, existential risk, ai safety}
	}
}

\usepackage{times}
\usepackage{url}
\usepackage{lipsum}
\usepackage{bbm}
\usepackage{booktabs}
\usepackage{tabularx}
\usepackage{makecell}
\newcolumntype{Y}{>{\centering\arraybackslash}X}
\newcolumntype{s}{>{\hsize=.3\hsize}Y}
\newcolumntype{t}{>{\hsize=.7\hsize}X}
\newcolumntype{b}{X}
\newcolumntype{u}{>{\hsize=0.8\hsize}Y}
\usepackage{wrapfig}
\usepackage{graphicx}
\usepackage{subcaption}
\usepackage{multirow}
\usepackage{todonotes}
\usepackage{amssymb}
\usepackage{amsfonts}
\usepackage{microtype}      %
\usepackage{cleveref}
\usepackage{titling} %
\usepackage[most]{tcolorbox}

\usepackage{spverbatim}

\title{
\vspace{-25pt}
\rule[0.4cm]{\textwidth}{2pt}
{\bf X-Risk Analysis for AI Research}
\rule{\textwidth}{2pt} 
}
\date{}

\author{\textbf{Dan Hendrycks}\\
UC Berkeley\\
\and
\textbf{Mantas Mazeika}\\
UIUC
}

\usepackage{arydshln}

\newcommand{\reviewer}[3]{
	\expandafter\newcommand\csname #1\endcsname[1]{
		\textcolor{#3}{[#2: ##1]}
	}
}
\definecolor{neonpurple}{rgb}{0.3,0,1}
\reviewer{dan}{Dan}{neonpurple}

\AtBeginBibliography{\small}

\begin{document}

\maketitle

\vspace*{-35pt}
\begin{abstract}
\normalsize
Artificial intelligence (AI) has the potential to greatly improve society, but as with any powerful technology, it comes with heightened risks and responsibilities. Current AI research lacks a systematic discussion of how to manage long-tail risks from AI systems, including speculative long-term risks. Keeping in mind the potential benefits of AI, there is some concern that building ever more intelligent and powerful AI systems could eventually result in systems that are more powerful than us; some say this is like playing with fire and speculate that this could create existential risks (x-risks). To add precision and ground these discussions, we provide a guide for how to analyze AI x-risk, which consists of three parts: First, we review how systems can be made safer today, drawing on time-tested concepts from hazard analysis and systems safety that have been designed to steer large processes in safer directions. Next, we discuss strategies for having long-term impacts on the safety of future systems. Finally, we discuss a crucial concept in making AI systems safer by improving the balance between safety and general capabilities. We hope this document and the presented concepts and tools serve as a useful guide for understanding how to analyze AI x-risk.\looseness=-1
\end{abstract}

\def\impacttext{In this section, please analyze how this work shapes the process that will lead to advanced AI systems and how it steers the process in a safer direction.}

\def\qone{\textbf{Overview.} How is this work intended to reduce existential risks from advanced AI systems?}

\def\qtwo{\textbf{Direct Effects.} If this work directly reduces existential risks, what are the main hazards, vulnerabilities, or failure modes that it directly affects?}
    
\def\qthree{\textbf{Diffuse Effects.} If this work reduces existential risks indirectly or diffusely, what are the main contributing factors that it affects?}

\def\qfour{\textbf{What's at Stake?} What is a future scenario in which this research direction could prevent the sudden, large-scale loss of life? If not applicable, what is a future scenario in which this research direction could be highly beneficial?}

\def\qfive{\textbf{Result Fragility.} Do the findings rest on strong theoretical assumptions; are they not demonstrated using leading-edge tasks or models; or are the findings highly sensitive to hyperparameters?}

\def\qsix{\textbf{Problem Difficulty.} Is it implausible that any practical system could ever markedly outperform humans at this task?} %

\def\qseven{\textbf{Human Unreliability.} Does this approach strongly depend on handcrafted features, expert supervision, or human reliability?}

\def\qeight{\textbf{Competitive Pressures.} Does work towards this approach strongly trade off against raw intelligence, other general capabilities, or economic utility?}

\def\qnine{\textbf{Overview.} How does this improve safety more than it improves general capabilities?}

\def\qten{\textbf{Red Teaming.} What is a way in which this hastens general capabilities or the onset of x-risks?}

\def\qeleven{\textbf{General Tasks.} Does this work advance progress on tasks that have been previously considered the subject of usual capabilities research?}

\def\qtwelve{\textbf{General Goals.} Does this improve or facilitate research towards general prediction, classification, state estimation, efficiency, scalability, generation, data compression, executing clear instructions, helpfulness, informativeness, reasoning, planning, researching, optimization, (self-)supervised learning, sequential decision making, recursive self-improvement, open-ended goals, models accessing the Internet, or similar capabilities?}

\def\qthirteen{\textbf{Correlation With General Aptitude.} Is the analyzed capability known to be highly predicted by general cognitive ability or educational attainment?}

\def\qfourteen{\textbf{Safety via Capabilities.} Does this advance safety along with, or as a consequence of, advancing other capabilities or the study of AI?}

\def\qfifteen{\textbf{Other.} What clarifications or uncertainties about this work and x-risk are worth mentioning?}

\def\externalitiestext{In this section, please analyze how this work relates to general capabilities and how it affects the balance between safety and hazards from general capabilities.}

\section{Introduction}

Artificial intelligence (AI) has opened up new frontiers in science and technology. Recent advances in AI research have demonstrated the potential for transformative impacts across many pursuits, including biology \cite{tunyasuvunakool2021highly}, mathematics \cite{polu2020generative}, visual art \cite{ramesh2022hierarchical}, coding \cite{chen2021evaluating}, and general game playing \cite{schrittwieser2020mastering}. By amplifying and extending the intelligence of humans, AI is a uniquely powerful technology with high upsides. However, as with any powerful technology, it comes with heightened risks and responsibilities. To bring about a better future, we have to actively steer AI in a beneficial direction and engage in proactive risk management.

Substantial effort is already directed at improving the safety and beneficence of current AI systems. From deepfake detection to autonomous vehicle reliability, researchers actively study how to handle current AI risks and take these risks very seriously. However, current risks are not the only ones that require attention. In the intelligence and defense communities, it is common to also anticipate future risks which are not yet present but could eventually become important. Additionally, as the COVID-19 pandemic demonstrates, tail risks that are rare yet severe should not be ignored \cite{taleb2020statistical}. Proactiveness and preparedness are highly valuable, even for low-probability novel tail risks, and scientists would be remiss not to contemplate or analyze tail risks from AI. Preparing for tail events is not overly pessimistic, but rather prudent.

Some argue that tail risks from future AI systems could be unusually high, in some cases even constituting an existential risk (x-risk)---one that could curtail humanity's long-term potential \cite{ord2020precipice}. Views on this topic fall across the spectrum. However, it is clear that building continually stronger AI systems at least amplifies existing risks, such as weaponization and disinformation at scale. Assuming continued progress, there is a distinct possibility of eventually building AIs that exceed human intelligence, which could usher in a new age of innovation but also create many new risks. While x-risk from AI is primarily future-oriented and often thought low probability, with some current estimates in the single digits over the next century \cite{carlsmith_power_seeking, grace2018will}, there is still substantial value in proactively gaining clarity on the risks and taking the anticipated hazards seriously.

Much research on AI safety is motivated by reducing x-risk \cite{hendrycks2021unsolved}. However, the literature currently lacks a grounded discussion of risk and tends to rely on a form of inchoate hazard analysis. We address this gap by providing a guide that introduces new concepts to understand and analyze AI x-risk.

In the main paper, we build up to discussing how to make strong AI systems safer by covering three key topics: how to make \textit{systems} safer, how to make \textit{future} systems safer, and finally how to make future \textit{AI} systems safer.
Specifically, in the first section we provide an overview of concepts from contemporary risk analysis and systems safety. These concepts have withstood the test of time across dozens of industries to enable the safe operation of diverse, high-risk complex systems without catastrophes \cite{leveson2016engineering, perrow1999normal}. Second, armed with a robust understanding of risk, we examine how today’s research can have a long-term impact on the development of safe AI, even though the future is far away and uncertain. Third, we discuss how na\"ive attempts to advance AI safety can backfire. To avoid this unintended consequence, we conclude the main paper by discussing how to improve the overall safety of AI systems by improving the \textit{balance} between safety and capabilities.

To further orient new AI x-risk researchers, we provide auxiliary background materials in the appendix.
In \Cref{sec:hazards}, we expand our discussion by elaborating on speculative hazards and failure modes that are commonplace concepts in AI x-risk discussions. In \Cref{app:unsolved}, we then describe concrete empirical research directions that aim to address the aforementioned hazards and failure modes. This culminates in X-Risk Sheets (\Cref{app:x-risk-sheet}), a new risk analysis tool to help researchers perform x-risk analysis of their safety research papers.\looseness=-1 %

We hope this document serves as a guide to safety researchers by clarifying how to analyze x-risks from AI systems, and helps stakeholders and interested parties with evaluating and assessing x-risk research contributions. Even though these risks may be low-probability and future-oriented, we should take them seriously and start building in safety early.

\section{Background AI Risk Concepts}\label{sec:reward_bias}

\subsection{General Risk Analysis}
To help researchers improve the safety of future AI systems, we provide a basic vocabulary and overview of concepts from general risk analysis. As with risks in many safety-critical systems, risks from strong AI can be better understood and managed with these concepts and tools, which have withstood the test of time across dozens of industries and applications. In particular, we cover basic terminology, discuss a risk decomposition, provide a precise model of reliability, describe safe design principles, and discuss the contemporary systems view of safety. Throughout this guide, at the end of each section we provide a concrete example in which we apply these concepts to analyze an AI research direction.

\paragraph{Definitions.}
A \emph{Hazard} is a source of danger with the potential to harm \cite{blanchard2008guide, leveson2016engineering}.
An \emph{Inherent Hazard} is a hazard that is inherently or characteristically posed by a system, such as hazardous materials in a chemical plant \cite{inherent}.
A \emph{Systemic Hazard} is a hazard from the broader sociotechnical system or social factors such as safety culture and management.
\emph{Exposure} is the extent to which elements (\textit{e.g.}, people, property, systems) are subjected or exposed to hazards.
\emph{Vulnerability} indicates susceptibility to the damaging effects of hazards, or a factor or process that increases susceptibility to the damaging effects of hazards.
A \emph{Threat} is a hazard with intent to exploit a vulnerability.
A \emph{Failure Mode} is a particular way a system might fail.
A \emph{Tail Risk} is a low-probability risk that can carry large consequences.
For completeness, an \emph{Existential Risk} or \emph{X-Risk} is a risk that can permanently curtail humanity's long-term potential \cite{ord2020precipice,bostrom2011global}.

\paragraph{Risk Equation.}
A decomposition of a risk from a given hazard can be captured by the notional equation
$\text{Risk} = \text{Hazard} \times \text{Exposure} \times \text{Vulnerability}$, where ``Hazard'' means hazard severity and prevalence, and ``$\times$'' merely indicates interaction. These specify risks from a particular hazard, and they can be aggregated as an expected risk by weighting with hazard probabilities. To illustrate this decomposition, the risk of chemical leaks from an industrial plant can be reduced by using less dangerous chemicals (reducing the hazard), building the plant far from populated areas (reducing exposure), and providing workers with protective clothing (reducing vulnerability).
Similarly, the risk of being in a car crash can be reduced by driving slower (reducing the hazard), driving on less dangerous roads (reducing exposure), or wearing a seatbelt (reducing vulnerability). In cybersecurity, the risk of a data leak from third-party vendors can be reduced by working with more trusted vendors (reducing the hazard), reducing vendor access to rooms where sensitive data is stored (reducing exposure), or by encrypting data so unauthorized vendors cannot interpret exfiltrated data (reducing vulnerability).\looseness=-1

The risk equation can be extended as: 
$\text{Risk} = \text{Hazard} \times \text{Exposure} \times \text{Vulnerability} \, / \, \text{Ability to Cope}$ to adjust for the ability to cope with or recover from accidents. This is relevant to risks from AI, because if we lose control of a strong AI system, our ability to cope may be zero. Likewise, by definition, x-risks are permanent, so this equation shows the risk of such events is limitlessly great.\looseness=-1

\begin{wrapfigure}{R}{0.45\textwidth}
	\vspace{-18pt}
	\centering
	\includegraphics[width=0.43\textwidth]{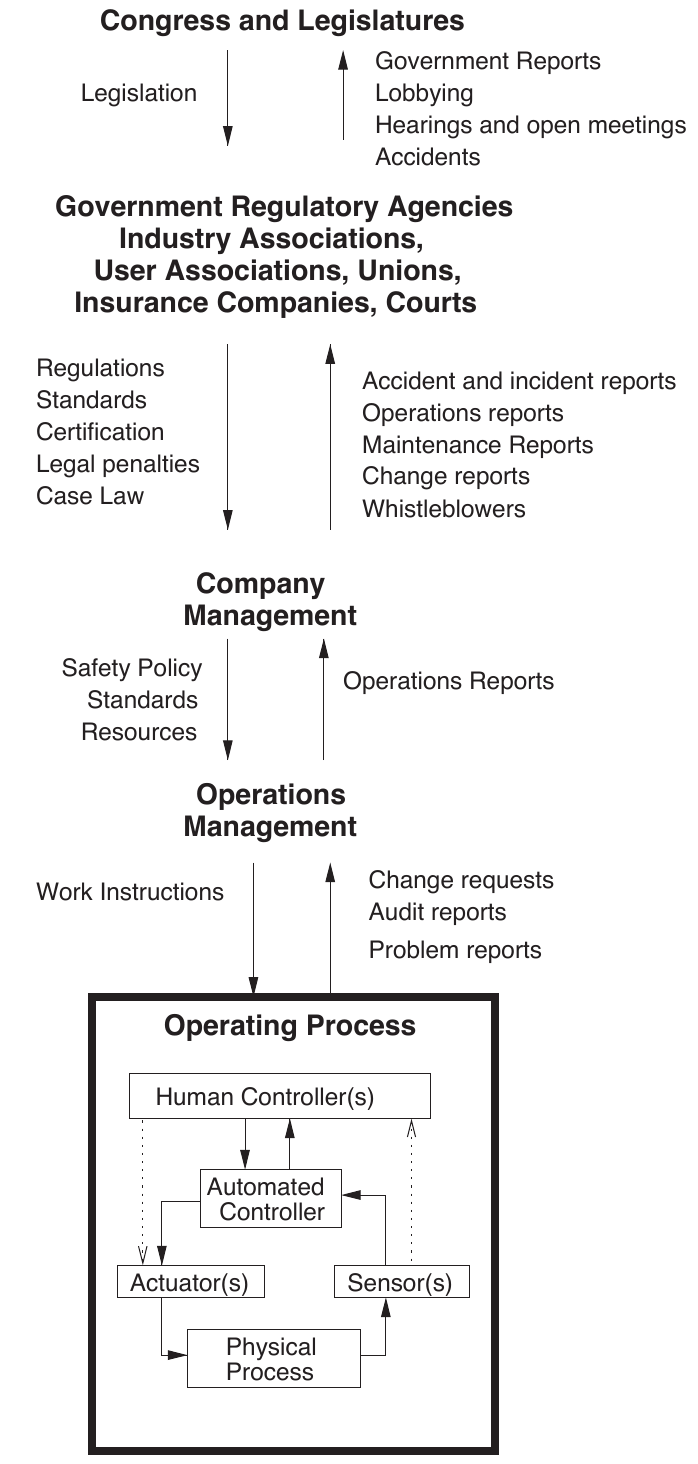}
	\caption{A simplified sociotechnical operational control structure (adapted from \cite{leveson2016engineering}). Systemic factors such as monitoring and safety culture are crucial for maintaining safe operations.}
	\label{fig:sociotechnical}
	\vspace{-15pt}
\end{wrapfigure}

\paragraph{Nines of Reliability.}
A helpful model for understanding reliability is the ``Nines of Reliability'' \cite{tao}. An event with a success probability $p$ has $k$ nines of reliability, where $k$ is defined as $k = - \log_{10}(1 - p)$. Hence if $p= 90\%, k=1$, and if $p=99.99\%, k=4$. For real-world systems, it is impossible to reach total reliability and zero risk due to the presence of adversaries, long-tails, emergence, and unknown unknowns. However, one can continue increasing nines of reliability to approach ideal safety.

In many settings, there is a sufficient level of reliability past which risks are acceptable. However, this is not the case for existential risks, because they threaten permanent failure and thus cannot be tolerated even once. This qualitative distinction between existential risk and normal risk highlights the importance of continually increasing nines of reliability for systems that create existential risk. Simplistically supposing a Poisson process for an existential catastrophe, adding $1$ nine of reliability corresponds to a $10\times$ reduction in the probability of permanent failure, resulting in a $10\times$ longer civilizational lifetime on average. Thus, increasing nines of reliability produces high value and does not suffer steeply diminishing returns.

\paragraph{Safe Design Principles.}
Safety-critical systems across different industries have several design principles in common. These principles could also make AI systems safer. One such design principle is \emph{redundancy}, which describes using similar components to remove single points of failure. \emph{Defense in depth} layers multiple defenses so that weaknesses or failures of one defense layer can be prevented by another defense layer. \emph{Transparency} improves our ability to understand and reason about systems. The \emph{principle of least privilege} means agents should have the minimum permissions and power necessary to accomplish tasks.
\emph{Loose coupling} of components makes a rapid cascade of failures less likely, and it increases controllability and time for readjustment \cite{perrow1999normal}. Imposing a \emph{separation of duties} implies no single agent has the ability to solely control or misuse the system on their own.
\emph{Fail-safes} are features that help systems fail gracefully \cite{adkins2020building}.

\paragraph{Systemic Factors.}
It is simplistic to require that all work demonstrate that it reduces risks directly.
Contemporary hazard analysis takes a systems view to analyze hazards since exclusively analyzing failure modes directly has well-known blind spots. 
Older analysis tools often assume that a ``root cause'' triggers a sequence of events that directly and ultimately cause a failure, but such models only capture linear causality.
Modern systems are replete with nonlinear causality, including feedback loops, multiple causes, circular causation, self-reinforcing processes, butterfly effects, microscale-macroscale dynamics, emergent properties, and so on. 
Requiring that researchers establish a direct link from their work to a failure mode erroneously and implicitly requires stories with linear causality and excludes nonlinear, remote, or indirect causes \cite{Leveson2009MovingBN}.
Backward chaining from the failure mode to the ``root cause'' and representing failures as an endpoint in a chain of events unfortunately leaves out many crucial causal factors and processes.
Rather than determine what event or component is the ``root cause'' ultimately responsible for a failure, in complex systems it is more fruitful to ask how various factors contributed to a failure \cite{cook}. In short, safety is far from just a matter of directly addressing failure modes \cite{perrow1999normal,hros}; safety is an emergent property \cite{leveson2016engineering} of a complex sociotechnical system comprised of many interacting, interdependent factors that can directly or indirectly cause system failures.

Researchers aware of contemporary hazard analysis could discuss how their work bears on these crucial indirect causes or diffuse contributing factors, even if their work does not fix a specific failure mode directly. We now describe many of these contributing factors.
\emph{Safety culture} describes attitudes and beliefs of system creators towards safety.
Safety culture is ``the most important to fix if we want to prevent future accidents'' \cite{talk}.
A separate factor is \emph{safety feature costs}; reducing these costs makes future system designers more likely to include additional safety features. The aforementioned \emph{safe design principles} can diffusely improve safety in numerous respects.
\emph{Improved monitoring tools} can reduce the probability of alarm fatigue and operators ignoring warning signs. Similarly, a reduction in \emph{inspection and preventative maintenance} can make failures more likely.
\emph{Safety team resources} is a critical factor, which consists of whether a safety team exists, headcount, the amount of allotted compute, dataset collection budgets, and so on.
A field's \emph{incentive structure} is an additional factor, such as whether people are rewarded for improving safety, even if it does not advance capabilities. \emph{Leadership epistemics} describes to the level of awareness, prudence, or wisdom of leaders or the quality of an organization's collective intelligence.
\emph{Productivity pressures} and \emph{competition pressures} can lead teams to cut corners on safety, ignore troubling signs, or race to the bottom.
Finally, \emph{social pressures} and \emph{rules and regulations} often help retroactively address failure modes and incentivize safer behavior. An example sociotechnical control structure is in \Cref{fig:sociotechnical}, illustrating the complexity of modern sociotechnical systems and how various systemic factors influence safety.

\paragraph{Application: Anomaly Detection.} To help concretize our discussion, we apply the various concepts in this section to anomaly detection. Anomaly detection helps identify \emph{hazards} such as novel \emph{failure modes}, and it helps reduce an operator's \emph{exposure} to hazards. Anomaly detection can increase the \emph{nines of reliability} of a system by detecting unusual system behavior before the system drifts into a more hazardous state. Anomaly detection helps provide \emph{defense in depth} since it can be layered with other safety measures, and it can trigger a \emph{fail-safe} when models encounter unfamiliar, highly uncertain situations. For sociotechnical systems, improved anomaly detectors can reduce the prevalence of \textit{alarm fatigue}, automate aspects of \textit{problem reports and change reports}, reduce \textit{safety feature costs}, and make \textit{inspection and preventative maintenance} less costly.

\subsection{AI Risk Analysis}

Existential risks from strong AI can be better understood using tools from general risk analysis. Here, we discuss additional considerations and analysis tools specific to AI risk.

\paragraph{Safety Research Decomposition.}
Research on AI safety can be separated into four distinct areas: robustness, monitoring, alignment, and systemic safety. \emph{Robustness} research enables withstanding hazards, including adversaries, unusual situations, and Black Swans \cite{taleb2020statistical}. \emph{Monitoring} research enables identifying hazards, including malicious use, hidden model functionality, and emergent goals and behaviors. \emph{Alignment} research seeks to make AI systems less hazardous by focusing on hazards such as power-seeking tendencies, dishonesty, or hazardous goals. \emph{Systemic Safety} research seeks to reduce system-level risks, such as malicious applications of AI and poor epistemics. These four research areas constitute high-level safety research priorities that can provide defense in depth against AI risks \cite{hendrycks2021unsolved}.

We can view these areas of AI safety research as tackling different components of the risk equation for a given hazard, $\text{Risk} = \text{Vulnerability} \times \text{Exposure} \times \text{Hazard}$. Robustness reduces vulnerability, monitoring reduces exposure to hazards, alignment reduces the prevalence and severity of inherent model hazards, and systemic safety reduces systemic risks by decreasing vulnerability, exposure, and hazard variables. The monitoring and systemic safety research areas acknowledge that hazards are neither isolated nor independent, as safety is an emergent property that requires improving direct as well as diffuse safety factors.

Alternatively, a large share of safety research could be categorized in one of these three areas: AI security, transparency, and machine ethics. \emph{AI Security} aims to make models cope in the face of adversaries. \emph{Transparency} aims to help humans reason about and understand AI systems. \emph{Machine Ethics} aims to create artificial agents that behave ethically, such as by not causing wanton harm.

\paragraph{Scope Levels.}
Drawing from \cite{DoD,hendrycks2021unsolved}, risks from strong AI can be separated into three scopes. First, \emph{AI System Risks} concern the ability of an individual AI system to operate safely. Examples of AI system risks include anomalies, adversaries, and emergent functionality. \emph{Operational Risks} concern the ability of an organization to safely operate an AI system during deployment. Examples of operational risks include alarm fatigue, model theft, competitive pressures that undervalue safety, and lack of safety culture. \emph{Institutional and Societal Risks} concern the ability of global society or institutions that decisively affect AI systems to operate in an efficient, informed, and prudent way. Examples of institutional and societal risks include an AI arms race, incentives for creating AI weapons or using AI to create weapons.

\begin{figure}
    \vspace{-40pt}
    \centering
    \includegraphics[width=\textwidth]{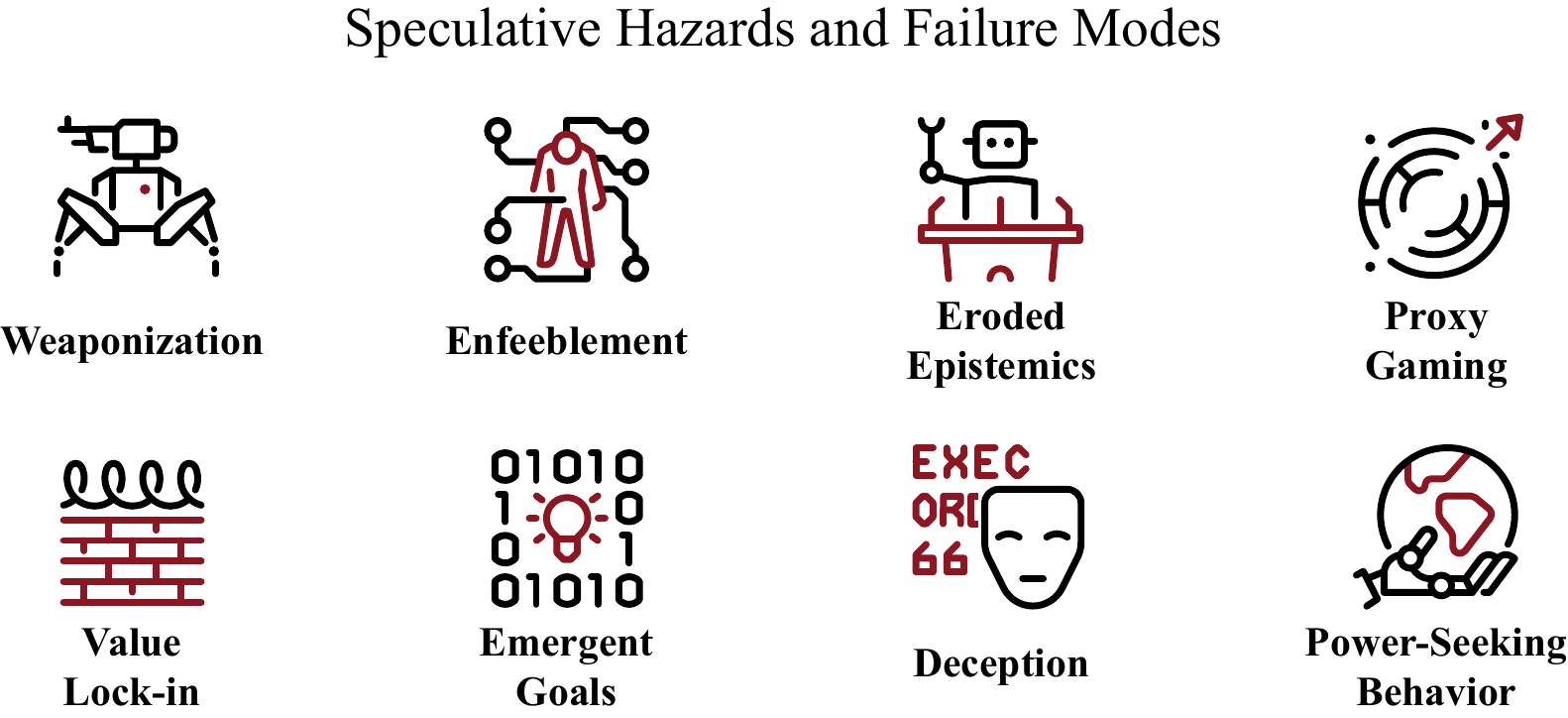}
    \caption{Speculative concerns about future AI systems that empirical research can make less likely.}
    \label{fig:hazards}
    \vspace{-15pt}
\end{figure}

\paragraph{Speculative Hazards and Failure Modes.}
Numerous \textit{speculative} hazards and failure modes contribute to existential risk from strong AI. \emph{Weaponization} is common for high-impact technologies. Malicious actors could repurpose AI to be highly destructive, and this could be an on-ramp to other x-risks; even deep RL methods and ML-based drug discovery have been successful in pushing the boundaries of aerial combat and chemical weapons \cite{dogfight,Urbina2022DualUO}, respectively. \emph{Enfeeblement} can occur if know-how erodes by delegating increasingly many important functions to machines; in this situation, humanity loses the ability to self-govern and becomes completely dependent on machines, not unlike scenarios in the film WALL-E. Similarly, \emph{eroded epistemics} would mean that humanity could have a reduction in rationality due to a deluge of misinformation or highly persuasive, manipulative AI systems.
\emph{Proxy misspecification} is hazardous because strong AI systems could over-optimize and game faulty objectives, which could mean systems aggressively pursue goals and create a world that is distinct from what humans value. \emph{Value lock-in} could occur when our technology perpetuates the values of a particular powerful group, or it could occur when groups get stuck in a poor equilibrium that is robust to attempts to get unstuck. \emph{Emergent functionality} could be hazardous because models demonstrate unexpected, qualitatively different behavior as they become more competent \cite{Ganguli2022PredictabilityAS,grokking}, so a loss of control becomes more likely when new capabilities or goals spontaneously emerge. \emph{Deception} is commonly incentivized, and smarter agents are more capable of succeeding at deception; we can be less sure of our models if we fail to find a way to make them assert only what they hold to be true. \emph{Power-seeking behavior} in AI is a concern because power helps agents pursue their goals more effectively \cite{turner2021optimal}, and there are strong incentives to create agents that can accomplish a broad set of goals; therefore, agents tasked with accomplishing many goals have instrumental incentives to acquire power, but this could make them harder to control \cite{carlsmith_power_seeking}. These concepts are visualized in \Cref{fig:hazards}, and we extend this discussion in \Cref{sec:hazards}. %

\paragraph{Application: Anomaly Detection.} To help concretize our discussion, we apply the various concepts in this section to anomaly detection. We first note that anomaly detection is a core function of \textit{monitoring}. In the short-term, anomaly detection reduces \textit{AI system risks} and \textit{operational risks}. It can help detect when \textit{misspecified proxies} are being overoptimized or gamed. It can also help detect misuse such as \textit{weaponization} research or \textit{emergent functionality}, and in the future it could possibly help detect AI \textit{deception}.

\section{Long-Term Impact Strategies}
While we have discussed important concepts and principles for making systems safer, how can we make strong AI systems safer given that they are in the future?
More generally, how can one affect the future in a positive direction, given that it is far away and uncertain?
People influence future decades in a variety of ways, including furthering their own education, saving for retirement decades in advance, raising children in a safe environment, and so on. Collectively, humans can also improve community norms, enshrine new rights, and counteract anticipated environmental catastrophes. Thus, while all details of the future are not yet known, there are successful strategies for generally improving long-term outcomes without full knowledge of the future.
Likewise, even though researchers do not have access to strong AI, they can perform research to reliably help improve long-term outcomes. In this section we discuss how empirical researchers can help shape the processes that will eventually lead to strong AI systems, and steer them in a safer direction. In particular, researchers can improve our understanding of the problem, improve safety culture, build in safety early, increase the cost of adversarial behavior, and prepare tools and ideas for use in times of crisis.\looseness=-1

\paragraph{Improve Field Understanding and Safety Culture.}
Performing high-quality research can influence our field's understanding and set precedents. High-quality datasets and benchmarks concretize research goals, make them tractable, and spur large community research efforts. Other research can help identify infeasible solutions or dead ends, or set new directions by identifying new hazards and vulnerabilities. At the same time, safety concerns can become normalized and precedents can become time-tested and standardized. These second-order effects are not secondary considerations but are integral to any successful effort toward risk reduction.\looseness=-1

\paragraph{Build In Safety Early.}
Many early Internet protocols were not designed with safety and security in mind. Since safety and security features were not built in early, the Internet remains far less secure than it could have been, and we continue to pay large continuing costs as a consequence.
Aggregating findings from the development of multiple technologies, a report for the Department of Defense \cite{Frola1984SystemSI} estimates that approximately $75\%$ of safety-critical decisions occur early in a system's development. Consequently, working on safety early can have \textit{founder effects}.
Moreover, incorporating safety features late in the design process is at times simply infeasible, leaving system developers no choice but to deploy without important safety features. In less extreme situations, retrofitting safety features near the end of a system's development imposes higher costs compared to integrating safety features earlier.

\paragraph{Improve Cost-Benefit Variables.}
Future decision-makers directing AI system development will need to decide which and how many safety features to include. These decisions will be influenced by a cost-benefit analysis. Researchers can decrease the capabilities costs of safety features and increase their benefits by doing research today.
In addition to decreasing safety feature costs, researchers can also increase the cost of undesirable adversarial behavior. Today and in the future, adversarial humans and adversarial artificial agents \cite{Gleave2020AdversarialPA} attempt to exploit vulnerabilities in machine learning systems. Removing vulnerabilities increases the cost necessary to mount an attack. Increasing costs makes adversaries less likely to attack, makes their attacks less potent, and can impel them to behave better.

Driving up the cost of adversarial behavior is a long-term strategy, since it can be applied to safeguard against powerful adversaries including hypothetical strong AI optimizers. For example, the military and information assurance communities face powerful adversaries and often work to increase the cost of adversarial behavior. In this way, cost-benefit analysis can comport with security from worst-case situations and adversarial forces. Additionally, this framework is more realistic than finding perfect safety solutions, as increasing costs to undesirable behavior recognizes that risk cannot entirely be eliminated. In summary, we can progressively reduce vulnerabilities in future AI systems to better defend against future adversaries.

\paragraph{Prepare for Crises.}
In times of crisis, decisive decisions must be made. The decisions made during such a highly impactful, highly variable period can result in a turning point towards a better outcome.
For this reason, a Nobel Prize economist \cite{Friedman} wrote, ``Only a crisis - actual or perceived - produces real change. When that crisis occurs, the actions that are taken depend on the ideas that are lying around. That, I believe, is our basic function: to develop alternatives..., to keep them alive and available until the politically impossible becomes the politically inevitable.''
Similarly, Benjamin Franklin wrote that ``an ounce of prevention is worth a pound of cure.''
The upshot of both of these views is that proactively creating and refining safety methods can be highly influential. Work today can influence which trajectory is selected during a crisis, or it can have an outsized impact in reducing a catastrophe's severity.

\paragraph{Prioritize by Importance, Neglectedness, and Tractability on the Margin.}
Since there are many problems to work on, researchers will need to prioritize. Clearly important problems are useful to work on, but if the problem is crowded, a researcher’s expected marginal impact is lower. Furthermore, if researchers can hardly make progress on a problem, the expected marginal impact is again lower.

Three factors that affect prioritization include importance, neglectedness, and tractability.
By ``importance,'' we mean the amount of risk reduced, assuming substantial progress.
A problem is more important if it is an x-risk and greatly influences many, not just one, existential failure modes. By ``neglectedness,'' we mean the extent to which a problem is relatively underexplored. A problem is more likely to be neglected if it is related to human values that are neglected by the maximization of economic preferences (e.g., meaning, equality, etc.), is out of the span of most researchers’ skillsets, primarily helps address rare but highly consequential Black Swans, addresses diffuse externalities, primarily addresses far-future concerns, or is not thought respectable or serious. Finally, by ``tractability,'' we mean the amount of progress that would likely be made on the problem assuming additional resources. A problem is more likely to be tractable if it has been concretized by measurable benchmarks and if researchers are demonstrating progress on those benchmarks.

\paragraph{Application: Anomaly Detection.}
To help concretize our discussion, we apply the various concepts in this section to anomaly detection. Anomaly detection is a concrete measurable problem, which can improve \textit{safety culture} and the \textit{field's understanding} of hazards such as unknown unknowns. As anomaly detection for deep learning began several years ago, there has been an attempt to \textit{build in safety early}. Consequently, this has led to more mature anomaly detection techniques than would have existed otherwise, thereby improving the \textit{benefit variables} of this safety feature. Consequently, in a future time of \textit{crisis} or a pivotal event, anomaly detection methods could be simple and reliable enough for inclusion in regulation. Last, anomaly detection's  \textit{importance} is high, and \textit{neglectedness} and \textit{tractability} are similar to other safety research avenues (see \Cref{app:unsolved}).\looseness=-1

\section{Safety-Capabilities Balance}

We discussed how to improve the safety of future systems in general. However, an additional concept needed to analyze future risks from AI in particular is the safety-capabilities balance, without which there has been much confusion about how to unmistakably reduce x-risks. First we discuss the association of safety and capabilities and distinguish the two. Then we discuss how well-intentioned pursuits towards safety can have unintended consequences, giving concrete examples of safety research advancing capabilities and vice versa.
To avoid future unintended consequences, we propose that researchers demonstrate that they improve the balance between safety and capabilities.\looseness=-1

As a preliminary, we note that ``general capabilties'' relates to concepts such as a model's accuracy on typical tasks, sequential decision making abilities in typical environments, reasoning abilities on typical problems, and so on. Due to the no free lunch theorem \cite{Wolpert1996TheLO}, we do not mean all mathematically definable tasks.\looseness=-1

\begin{wrapfigure}{R}{0.35\textwidth}
	\vspace{-10pt}
	\centering
	\includegraphics[width=0.35\textwidth]{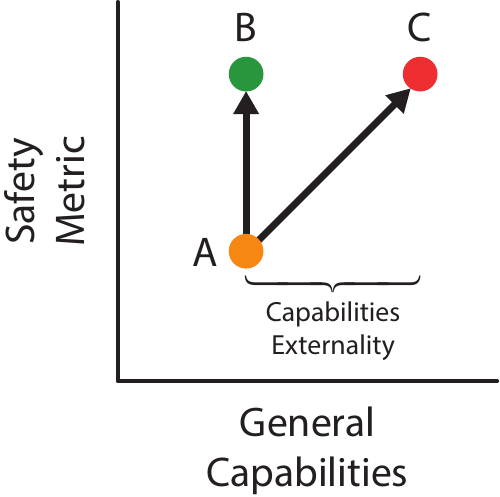}
	\caption{Model B is orthogonal to general capabilities and more clearly improves safety. Model C does not improve the safety-capabilities balance, and since greater capabilities can harm or help safety, its impact on overall safety is less clear than Model B.
	}
	\label{fig:externalities}
	\vspace{-10pt}
\end{wrapfigure}

\paragraph{Intelligence Can Help or Harm Safety.}
Models that are made more intelligent could more easily avoid failures and act more safely. At the same time, models of greater intelligence could more easily act destructively or be directed maliciously. Similarly, a strong AI could help us make wiser decisions and help us achieve a better future, but loss of control is also a possibility. Raw intelligence is a double-edged sword and is not inextricably bound to desirable behavior. For example, it is well-known that \emph{moral virtues} are distinct from \emph{intellectual virtues}. An agent that is knowledgeable, inquisitive, quick-witted, and rigorous is not necessarily honest, just, power-averse, or kind \cite{aristotle,hume,Armstrong2013GeneralPI}. Consequently we want our models to have more than just raw intelligence.\looseness=-1

\paragraph{Side Effects of Optimizing Safety Metrics.}
Attempts to endow models with more than raw intelligence can lead to unintended consequences. In particular, attempts to pursue safety agendas can sometimes hasten the onset of AI risks. For example, suppose that a safety goal is concretized through a safety metric, and a researcher tries to create a model that improves that safety metric. In \Cref{fig:externalities}, the researcher could improve on model A just by improving the safety metric (model B), or by advancing the safety metric and general capabilities simultaneously (model C). In the latter case, the researcher improved general capabilities and as a side-effect has increased model intelligence, which we have established has a mixed impact on safety. We call such increases \emph{capabilities externalities}, shown in \Cref{fig:externalities}. This is not to suggest capabilities are bad or good \textit{per se}---they can help or harm safety and will eventually be necessary for helping humanity reach its full potential.\looseness=-1

\paragraph{Examples of Capabilities $\to$ Safety Goals.} We now provide concrete examples to illustrate how safety and general capabilities are associated.
Self-supervised learning can increase accuracy and data efficiency, but it can also improve various safety goals in robustness and monitoring \cite{Hendrycks2019UsingSL}. Pretraining makes models more accurate and extensible, but it also improves various robustness and monitoring goals \cite{Hendrycks2019UsingPC}. 
Improving an agent's world model makes them more generally capable, but this also can make them less likely to spawn unintended consequences. Optimizers operating over longer time horizons will be able to accomplish more difficult goals, but this could also make models act more prudently and avoid taking irreversible actions.

\paragraph{Examples of Safety Goals $\to$ Capabilities.}
Some argue that a safety goal is modeling user preferences, but depending on the preferences modeled, this can have predictable capabilities externalities. Recommender, advertisement, search, and machine translation systems make use of human feedback and revealed preferences to improve their systems. Recent work on language models uses reinforcement learning to incorporate user preferences over a general suite of tasks, such as summarization, question-answering, and code generation \cite{gao2020dialogrpt,Ouyang2022TrainingLM,Bai2022TrainingAH}. Leveraging task preferences, often styled as ``human values,'' can amount to making models more generally intelligent, as users prefer smarter models. Rather than model task preferences, researchers could alternatively minimize capabilities externalities by modeling timeless human values such as normative factors \cite{structure} and intrinsic goods (e.g., pleasure, knowledge, friendship, and so on).\looseness=-1

Some argue that a safety goal is truthfulness, but making models more truthful can have predictable capabilities externalities.
Increasing truthfulness can consist of increasing accuracy, calibration, and honesty. Increasing standard accuracy clearly advances general capabilities, so researchers aiming to clearly improve safety would do well to work more specifically towards calibration and honesty.

\paragraph{Safety-Capabilities Ratio.}
As we have seen, improving safety metrics does not necessarily improve our overall safety.
Improving a safety metric can improve our safety, all else equal. However, often all else is not equal since capabilities are also improved, so our overall safety has not necessarily increased. Consequently, to move forward, safety researchers must perform a more holistic risk analysis that simultaneously reports safety metrics and capabilities externalities, so as to demonstrate a reduction in total risk. We suggest that researchers improve the \textit{balance} between safety and general capabilities or, so to speak, improve a safety-capabilities ratio. To be even more precautionary and have a less mixed effect on safety, we suggest that safety research aim to avoid general capabilities externalities. This is because safety research should consistently improve safety more than it would have been improved by default.

This is certainly not to suggest that safety research is at odds with capabilities research---the overall effects of increased capabilities on safety are simply mixed, as established earlier. While developing AI precautiously, it would be beneficial to avoid a counterproductive ``safety vs. capabilities'' framing. Rather, capabilities researchers should increasingly focus on the potential benefits from AI, and safety researchers should focus on minimizing any potential tail risks. This process would function best if done collaboratively rather than adversarially, in much the same way information security software engineers collaborate with other software engineers. While other researchers advance general capabilities, safety researchers can differentially \cite{Bostrom2002ExistentialRA} improve safety by improving the safety-capabilities ratio.

\begin{wrapfigure}{R}{0.5\textwidth}
	\vspace{-17pt}
	\centering
	\includegraphics[width=0.48\textwidth]{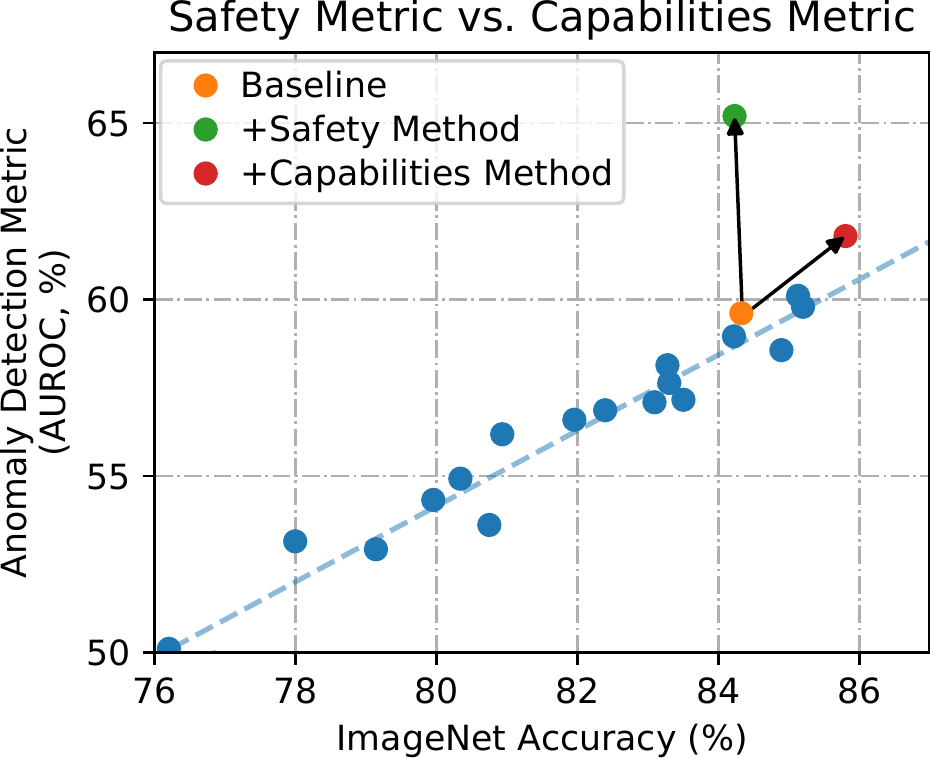}
	\caption{ImageNet model accuracies and their corresponding anomaly detection performance. The capabilities method does not make the model safer than it would be by default.}
	\label{fig:concrete}
	\vspace{-8pt}
\end{wrapfigure}

We now consider objections to this view. Some researchers may argue that we need to advance capabilities (e.g., reasoning, truth-finding, and contemplation capabilities) to study some long-term safety problems. This does not appear robustly beneficial for safety and does not seem necessary, as there are numerous neglected, important, and tractable existing research problems. In contrast,  advancing general capabilities is obviously not neglected. Additionally, safety researchers could rely on the rest of the community to improve upstream capabilities and then eventually use these capabilities to study safety-relevant problems. Next, research teams may argue that they have the purest motivations, so their organization must advance capabilities to race ahead of the competition and build strong AI as soon as possible. Even if a large advantage over the whole field can be reliably and predictably sustained, which is highly dubious, this is not necessarily a better way to reduce risks than channeling additional resources towards safety research.
Finally, some argue that work on safety will lead to a false perception of safety and cause models to be deployed earlier. Currently many companies clearly lack credible safety efforts (e.g., many companies lack a safety research team), but in the future the community should be on guard against a false sense of security, as is important in other industries.\looseness=-1 %

\paragraph{Practical Recommendations.}
To help researchers have a less mixed and clearer impact on safety, we suggest two steps. First, researchers should empirically measure the extent to which their method improves their safety goal or metric (e.g., anomaly detection AUROC, adversarial robustness accuracy, etc.); more concrete safety goals can be found in \cite{hendrycks2021unsolved} and in \Cref{app:unsolved}. Second, researchers should measure whether their method can be used to increase general capabilities by measuring its impact on correlates of general capabilities (e.g., reward in Atari, accuracy on ImageNet, etc.). With these values estimated, researchers can determine whether they differentially improved the balance between safety and capabilities. More precautious researchers can also note whether their improvement is approximately orthogonal to general capabilities and has minimal capabilities externalities. This is how empirical research claiming to differentially improve safety can demonstrate a differential safety improvement empirically.

\paragraph{Application: Anomaly Detection.}
To help concretize our discussion, we apply the various concepts in this section to anomaly detection. As shown in \Cref{fig:concrete}, anomaly detection safety measures are correlated with the accuracy of vanilla models, but differential progress is possible without simply increasing accuracy. A similar plot is in a previous research paper \cite{Hendrycks2021TheMF}. The plot shows that it is possible to improve anomaly detection without substantial \textit{capabilities externalities}, so work on anomaly detection can improve the \textit{safety-capabilities balance}.\looseness=-1

\section{Conclusion}
We provided a guiding document to help researchers understand and analyze AI x-risk. First, we reviewed general concepts for making systems safer, grounding our discussion in contemporary hazard analysis and systems safety. Next, we discussed how to influence the safety of future systems via several long-term impact strategies, showing how individual AI researchers can make a difference. Finally, we presented an important AI-specific consideration of improving the safety-capabilities balance. We hope our guide can clarify how researchers can reduce x-risk in the long term and steer the processes that lead to strong AI in a safer direction.\looseness=-1

\newpage

\section*{Acknowledgments}
We thank Thomas Woodside, Kevin Liu, Sidney Hough, Oliver Zhang, Steven Basart, Shudarshan Babu, Daniel McKee, Boxin Wang, Victor Gonzalez, Justis Mills, and Huichen Li for feedback.
DH is supported by the NSF GRFP Fellowship and an Open Philanthropy Project AI Fellowship.
\printbibliography

\appendix

\newpage
\section{An Expanded Discussion of Speculative Hazards and Failure Modes}\label{sec:hazards}
We continue our guide by providing an expanded discussion of the eight aforementioned speculative hazards and failure modes, namely weaponization, enfeeblement, eroded epistemics, proxy misspecification, value lock-in, emergent functionality, deception, and power-seeking behavior.

\begin{enumerate}[leftmargin=*]
\item \emph{Weaponization}: Some are concerned that weaponizing AI may be an onramp to more dangerous outcomes.
In recent years, deep RL algorithms can outperform humans at aerial combat \cite{dogfight}, AlphaFold has discovered new chemical weapons \cite{Urbina2022DualUO}, researchers have been developing AI systems for automated cyberattacks \cite{Buchanan2020Cyber,Cary2020DestructiveCO}, military leaders have discussed having AI systems have decisive control over nuclear silos \cite{c2}, and superpowers of the world have declined to sign agreements banning autonomous weapons. Additionally, an automated retaliation system accident could rapidly escalate and give rise to a major war.
Looking forward, we note that since the nation with the most intelligent AI systems could have a strategic advantage, it may be challenging for nations not to build increasingly powerful weaponized AI systems.

Even if AI alignment is solved and all superpowers agree not to build destructive AI technologies, rogue actors still could use AI to cause significant harm. Easy access to powerful AI systems increases the risk of unilateral, malicious usage. As with nuclear and biological weapons, only one irrational or malevolent actor is sufficient to unilaterally cause harm on a massive scale. Unlike previous weapons, stealing and widely proliferating powerful AI systems could just be a matter of copy and pasting.

\item \emph{Enfeeblement}: As AI systems encroach on human-level intelligence, more and more aspects of human labor will become faster and cheaper to accomplish with AI. As the world accelerates, organizations may voluntarily cede control to AI systems in order to keep up. This may cause humans to become economically irrelevant, and once AI automates aspects of many industries, it may be hard for displaced humans to reenter them. In this world, humans could have few incentives to gain knowledge or skills. These trends could lead to human enfeeblement and reduce human flourishing, leading to a world that is undesirable. Furthermore, along this trajectory, humans would have less control of the future.

\item \emph{Eroded epistemics}: States, parties, and organizations use technology to influence and convince others of their political beliefs, ideologies, and narratives. Strong AI may bring this use-case into a new era and enable personally customized disinformation campaigns at scale. Additionally, AI itself could generate highly persuasive arguments that invoke primal human responses and inflame crowds. Together these trends could undermine collective decision-making, radicalize individuals, derail moral progress, or erode consensus reality.

\item \emph{Proxy misspecification}: AI agents are directed by goals and objectives. Creating general-purpose objectives that capture human values could be challenging. As we have seen, easily measurable objectives such as watch time and click rates often trade off with our actual values, such as wellbeing \cite{kross2013facebook}. For instance, well-intentioned AI objectives have unexpectedly caused people to fall down conspiracy theory rabbit holes.
This demonstrates that organizations have deployed models with flawed objectives and that creating objectives which further human values is an unsolved problem.
Since goal-directed AI systems need measurable objectives, by default our systems may pursue simplified proxies of human values. The result could be suboptimal or even catastrophic if a sufficiently powerful AI successfully optimizes its flawed objective to an extreme degree.

\item \emph{Value lock-in}: Strong AI imbued with particular values may determine the values that are propagated into the future. Some argue that the exponentially increasing compute and data barriers to entry make AI a centralizing force. As time progresses, the most powerful AI systems may be designed by and available to fewer and fewer stakeholders. This may enable, for instance, regimes to enforce narrow values through pervasive surveillance and oppressive censorship. Overcoming such a regime could be unlikely, especially if we come to depend on it. Even if creators of these systems know their systems are self-serving or harmful to others, they may have incentives to reinforce their power and avoid distributing control. The active collaboration among many groups with varying goals may give rise to better goals \cite{sep-jury-theorems}, so locking in a small group's value system could curtail humanity's long-term potential.

\item \emph{Emergent functionality}: Capabilities and novel functionality can spontaneously emerge in today's AI systems \cite{Ganguli2022PredictabilityAS,grokking}, even though these capabilities were not anticipated by system designers. If we do not know what capabilities systems possess, systems become harder to control or safely deploy. Indeed, unintended latent capabilities may only be discovered during deployment. If any of these capabilities are hazardous, the effect may be irreversible.

New system objectives could also emerge. For complex adaptive systems, including many AI agents, goals such as self-preservation often emerge \cite{HadfieldMenell2017TheOG}. Goals can also undergo qualitative shifts through the emergence of \emph{intrasystem goals} \cite{gall2002systems,hendrycks2021unsolved}. In the future, agents may break down difficult long-term goals into smaller subgoals. However, breaking down goals can distort the objective, as the true objective may not be the sum of its parts. This distortion can result in misalignment. In more extreme cases, the intrasystem goals could be pursued at the expense of the overall goal. For example, many companies create intrasystem goals and have different specializing departments pursue these distinct subgoals. However, some departments, such as bureaucratic departments, can capture power and have the company pursue goals unlike its original goals. Even if we correctly specify our high-level objectives, systems may not operationally pursue our objectives \cite{Hubinger2019RisksFL}. This is another way in which systems could fail to optimize human values.

\item \emph{Deception}: Future AI systems could conceivably be deceptive not out of malice, but because deception can help agents achieve their goals. It may be more efficient to gain human approval through deception than to earn human approval legitimately. Deception also provides optionality: systems that have the capacity to be deceptive have strategic advantages over restricted, honest models. Strong AIs that can deceive humans could undermine human control.

AI systems could also have incentives to bypass monitors. Historically, individuals and organizations have had incentives to bypass monitors. For example, Volkswagen programmed their engines to reduce emissions only when being monitored. This allowed them to achieve performance gains while retaining purportedly low emissions. Future AI agents could similarly switch strategies when being monitored and take steps to obscure their deception from monitors. Once deceptive AI systems are cleared by their monitors or once such systems can overpower them, these systems could take a ``treacherous turn'' and irreversibly bypass human control.

\item \emph{Power-seeking behavior}:  Agents that have more power are better able to accomplish their goals. Therefore, it has been shown that agents have incentives to acquire and maintain power \cite{turner2021optimal}.
AIs that acquire substantial power can become especially dangerous if they are not aligned with human values. Power-seeking behavior can also incentivize systems to pretend to be aligned, collude with other AIs, overpower monitors, and so on. On this view, inventing machines that are more powerful than us is playing with fire. Building power-seeking AI is also incentivized because political leaders see the strategic advantage in having the most intelligent, most powerful AI systems. For example, Vladimir Putin has said ``Whoever becomes the leader in [AI] will become the ruler of the world.''

\end{enumerate}
\newpage
\section{Unsolved Problems in AI X-Risk}\label{app:unsolved}
In this section we describe empirical research directions towards reducing x-risk from AI.
We describe each problem, give its motivation, and suggest what late-stage quality work could look like. When describing potential advanced work, we hope that work developed is subject to the capabilities externalities constraints discussed earlier.

\subsection{Adversarial Robustness}
Adversarial examples demonstrate that optimizers can easily manipulate vulnerabilities in AI systems and cause them to make egregious mistakes. Adversarial vulnerabilities are long-standing weaknesses of AI models. While typical adversarial robustness is related to AI x-risk, future threat models are broader than today’s adversarial threat models. Since we are concerned about being robust to optimizers that cause models to make mistakes generally, we need not assume that optimizers are subject to small, specific $\ell_p$ distortion constraints, as their attacks could be unforeseen and even perceptible. We also need not assume that a human is in the loop and can check if an example is visibly distorted. In short, this area is about making AI systems robust to powerful optimizers that aim to induce specific system responses.

\paragraph{Motivation.} In the future, AI systems may pursue goals specified by other AI proxies. For example, an AI could encode a proxy for human values, and another AI system could be tasked with optimizing the score assigned by this proxy. The quality of an AI agent's actions would be judged by the AI proxy, and the agent would conform its conduct to receive high scores from the AI proxy. If the human value proxy instantiated by the AI is not robust to optimizers, then its vulnerabilities could be exploited, so this gameable proxy may not be fully safe to optimize. By improving the reliability of learned human value proxies, optimizers would have a harder time gaming these systems. If gaming becomes sufficiently difficult, the optimizer can be impelled to optimize the objective correctly. Separately, humans and systems will monitor for destructive behavior, and these monitoring systems need to be robust to adversaries.

\paragraph{What Advanced Research Could Look Like.}
Ideally, an adversarially robust system would make reliable decisions given adversarially constructed inputs, and it would be robust to adversaries with large attack budgets using unexpected novel attacks. Furthermore, it should detect adversarial behavior and adversarially optimized inputs. A hypothetical human value function should be as adversarially robust as possible so that it becomes safer to optimize. A hypothetical human value function that is fully adversarially robust should be safe to optimize.

\subsection{Anomaly Detection}
\paragraph{Problem Description.} This area is about detecting potential novel hazards such as unknown unknowns, unexpected rare events, or emergent phenomena. Anomaly detection can allow models to flag salient anomalies for human review or execute a conservative fallback policy.

\paragraph{Motivation.} There are numerous existentially relevant hazards that anomaly detection could possibly identify more reliably or identify earlier, including proxy gaming, rogue AI systems, deception from AI systems, trojan horse models (discussed below), malicious use \cite{Brundage2018TheMU}, early signs of dangerous novel technologies \cite{Bostrom2019TheVW}, and so on.

For example, anomaly detection could be used to detect emergent and unexpected AI goals.
As discussed earlier, it is difficult to make systems safe if we do not know what they can do or how they differ from previous models. New instrumental goals may emerge in AI systems, and these goals may be undesirable or pose x-risks (such as the goal for a system to preserve itself, deceive humans, or seek power). If we can detect that a model has a new undesirable capability or goal, we can better control our systems through this protective measure against emergent x-risks.

\paragraph{What Advanced Research Could Look Like.}
A successful anomaly detector could serve as an AI watchdog that could reliably detect and triage rogue AI threats. When the watchdog detects rogue AI agents, it should do so with substantial lead time. Anomaly detectors should also be able to straightforwardly create tripwires for AIs that are not yet considered safe. Furthermore, advanced anomaly detectors should be able to help detect ``black balls'', meaning ``a technology that invariably or by default destroys the civilization that invents it'' \cite{bostrom2019vulnerable}. Anomaly detectors should also be able to detect biological hazards, by having anomaly detectors continually scan hospitals for novel biological hazards.

\subsection{Interpretable Uncertainty}
\paragraph{Problem Description.} This area is about making model uncertainty more interpretable and calibrated by adding features such as confidence interval outputs, conditional probabilistic predictions specified with sentences, posterior calibration methods, and so on.

\paragraph{Motivation.} If operators ignore system uncertainties since the uncertainties cannot be relied upon or interpreted, then this would be a contributing factor that makes the overall system that monitors and operates AIs more hazardous. To draw a comparison to chemical plants, improving uncertainty expressiveness could be similar to ensuring that chemical system dials are calibrated. If dials are uncalibrated, humans may ignore the dials and thereby ignore warning signs, which increases the probability of accidents and catastrophe.

Furthermore, since many questions in normative ethics have yet to be resolved, human value proxies should incorporate moral uncertainty. If AI human values proxies have appropriate uncertainty, there is a reduced risk in an human value optimizer maximizing towards ends of dubious value.

\paragraph{What Advanced Research Could Look Like.}
Future models should be calibrated on inherently uncertain, chaotic, or computationally prohibitive questions that extend beyond existing human knowledge. Their uncertainty should be easily understood by humans, possibly by having models output structured probabilistic models (``event A will occur with 60\% probability assuming event B also occurs, and with 25\% probability if event B does not''). Moreover, given a lack of certainty in any one moral theory, AI models should accurately and interpretably represent this uncertainty in their human value proxies.

\subsection{Transparency}
\paragraph{Problem Description.} AI systems are becoming more complex and opaque. This area is about gaining clarity about the inner workings of AI models, and making models more understandable to humans.

\paragraph{Motivation.} Transparency tools could help unearth deception, mitigating risks from dishonest AI and treacherous turns. Transparency tools may also potentially be useful for identifying emergent capabilities. Moreover, transparency tools could help us better understand strong AI systems, which could help us more knowledgeably direct them and anticipate their failure modes.

\paragraph{What Advanced Research Could Look Like.} Successful transparency tools would allow a human to predict how a model would behave in various situations without testing it. These tools could be easily applied (ex ante and ex post) to unearth deception, emergent capabilities, and failure modes.

\subsection{Trojans}
\paragraph{Problem Description.} AI systems can contain ``trojan’’ hazards. Trojaned models behave typically in most situations, but when specific secret situations are met, they reliably misbehave. For example, an AI agent could behave normally, but when given a special secret instruction, it could execute a coherent and destructive sequence of actions \cite{Foley2022ExecuteO6}. In short, this area is about identifying hidden functionality embedded in models that could precipitate a treacherous turn.

\paragraph{Motivation.} The trojans literature has shown that it is possible for dangerous, surreptitious modes of behavior to exist within AIs as a result of model weight editing or data poisoning. Misaligned AI or external actors could hide malicious behavior, such that it abruptly emerges at a time of their choosing. Future planning agents could have special plans unbeknownst to model designers, which could include plans for a treacherous turn. For this reason AI trojans provide a microcosm for studying treacherous turns.

\paragraph{What Advanced Research Could Look Like.} Future trojan detection techniques could reliably detect if models have trojan functionality. Other trojan research could develop reverse-engineering tools that synthesize or reconstruct the triggering conditions for trojan functionality. When applied to sequential decision making agents, this could potentially allow us to unearth surreptitious plans.

\subsection{Honest AI}
\paragraph{Problem Description.} Honest AI involves determining what models hold to be true, perhaps by analyzing their internal representations \cite{burns2022honest}. It is also about creating models that only output what they hold to be true.

\paragraph{Motivation.} If it is within a model's capacity to be strategically deceptive---able to make statements that the model in some sense knows to be false in order to gain an advantage---then treacherous turn scenarios are more feasible. Models could deceive humans about their plans, and then execute a new plan after the time when humans cannot course-correct. Plans for a treacherous turn could be brought to light by detecting dishonesty, or models could be made inherently honest, allowing operators to query them about their true plans.

\paragraph{What Advanced Research Could Look Like.} Good techniques could be able to reliably detect when a model's representations are at odds with its outputs. Models could also be trained to avoid dishonesty and allow humans to correctly conclude that models are being honest with high levels of certainty.

\subsection{Power Aversion}
\paragraph{Problem Description.} This area is about incentivizing models to avoid power or avoid gaining more power than is necessary.

\paragraph{Motivation.} Strategic AIs tasked with accomplishing goals would have instrumental incentives to accrue and maintain power, as power helps agents more easily achieve their goals. Likewise, some humans would have incentives to build and deploy systems that acquire power, because such systems would be more useful. If power-seeking models are misaligned, they could permanently disempower humanity.

\paragraph{What Advanced Research Could Look Like.} Models could evaluate the power of other agents in the world to accurately identify particular systems that were attaining more power than necessary. They could also be used to directly apply a penalty to models so that they are disincentivized from seeking power. Before agents pursue a task, other models could predict the types of power \cite{French1959TheBO} and amount of power they require. Lastly, models might be developed which are intrinsically averse to seeking power despite the instrumental incentive to seek power.

\subsection{Moral Decision-Making}
\paragraph{Problem Description.} This area is about building models to understand ethical systems and steering models to behave ethically.

\paragraph{Motivation.} This line of work helps create actionable ethical objectives for systems to pursue. If strong AIs are given objectives that are poorly specified, they could pursue undesirable actions and behave unethically. If these strong AIs are sufficiently powerful, these misspecifications could lead the AIs to create a future that we would strongly dislike. Consequently, work in this direction helps us avoid proxy misspecification as well as value lock-in.

\paragraph{What Advanced Research Could Look Like.} High-functioning models should detect situations where moral principles apply, assess how to apply those moral principles, evaluate the moral worth of candidate actions, select and carry out actions appropriate for the context, monitor the success or failure of those actions, and adjust responses accordingly \cite{railton}.

Models should represent various purported intrinsic goods, including pleasure, autonomy, the exercise of reason, knowledge, friendship, love, and so on. Models should be able to distinguish between subtly different levels of these goods, and their value functions should not be vulnerable to optimizers. Models should be able to create pros and cons of actions with respect to each of these values, and brainstorm how changes to a given situation would increase or decrease the amount of a given intrinsic good. They should also be able to create superhuman forecasts of how an action might affect these values in the long-term (e.g., how studying rather than procastinating can reduce wellbeing in the short-term but be useful for wellbeing in the long-term). Models should also be able to represent more than just intrinsic goods, as they should also be able to represent constantly-updating legal systems and normative factors including special obligations (such as parental responsibility) and deontological constraints.

Another possible goal is to create an automated moral parliament \cite{Newberry2021ThePA}, a framework for making ethical decisions under moral and empirical uncertainty. Sub-agents could submit their decisions to an internal moral parliament, which would incorporate the ethical beliefs of multiple stakeholders in informing decisions about which actions should be taken. Using a moral parliament could reduce the probability that we are leaving out important normative factors by focusing on only one moral theory, and the inherent multifaceted, redundant, ensembling nature of a moral parliament would also contribute to making models less gameable. If a component of the moral parliament is uncertain about a judgment, it could request help from human stakeholders. The moral parliament might also be able to act more quickly to restrain rogue agents than a human could, and therefore act effectively in the fast-moving world that is likely to be induced by more capable AI.

\subsection{Value Clarification}
\paragraph{Problem Description.} This area is about building AI systems that can perform moral philosophy research. This research area should utilize existing capabilities and avoid advancing general research, truth-finding, or contemplation capabilities.

\paragraph{Motivation.} Just in the past few decades, peoples' moral attitudes have changed on numerous issues. It is unlikely humanity's moral development is complete, and it is possible there are ongoing moral catastrophes.

To address deficiencies in our moral systems, and to more rapidly and wisely address future moral quandaries that humanity will face, these research systems could help us reduce risks of value lock-in by improving our moral precedents earlier rather than later. If humanity does not take a ``long reflection'' \cite{ord2020precipice} to consider and refine its values after it develops strong AI, then the value systems lying around may be amplified and propagated into the future. Value clarification reduces risks from locked-in, deficient value systems. Additionally, value clarification can be understood as a way to reduce proxy misspecification, as it can allow values to be updated in light of new situations or evidence.

\paragraph{What Advanced Research Could Look Like.} Advanced Research in value clarification would be able to produce original insights in philosophy, such that models could make philosophical arguments or write seminal philosophy papers. Value clarification systems could also point out inconsistencies in existing ethical views, arguments, or systems.

\subsection{ML for Cyberdefense}
\paragraph{Problem Description.} This area is about using machine learning to improve defensive security, such as by improving malicious program detectors. This area focuses on research avenues that are clearly defensive and not easily repurposed into offensive techniques, such as threat detectors and not automated penetration testers.

\paragraph{Motivation.} We care about improving computer security defenses for three main reasons. First, strong AI may be stored on private computers, and these computers would need to be secured. If they are not secured, and if strong AIs can be made destructive easily, then dangerous AI systems could be exfiltrated and widely proliferated. Second, AI systems that are hackable are not safe, as they could be maliciously directed by hackers. Third, cyberattacks could take down national infrastructure including power grids \cite{Ottis2008AnalysisOT}, and large-scale, reliable, and automated cyberattacks could engender political turbulence and great power conflicts \cite{Buchanan2020Cyber}. Great power conflicts incentivize countries to search the darkest corners of technology to develop devastating weapons. This increases the probability of weaponized AI, power-seeking AI, and AI facilitating the development of other unprecedented weapons, all of which are x-risks. Using ML to improve defense systems by decreasing incentives for cyberwarfare makes these futures less likely.

\paragraph{What Advanced Research Could Look Like.} AI-based security systems could be used for better intrusion detection, firewall design, malware detection, and so on.

\subsection{ML for Improving Epistemics}
\paragraph{Problem Description.} This area is about using machine learning to improve the epistemics and decision-making of political leaders. This area is tentative; if it turns out to have difficult-to-avoid capabilities externalities, then it would be a less fruitful area for improving safety.

\paragraph{Motivation.} We care about improving decision-making among political leaders to reduce the chance of rash or possibly catastrophic decisions. These decision-making systems could be used in high-stakes situations where decision makers do not have much foresight, where passions are inflamed, and decisions must be made extremely quickly, perhaps based on gut reactions. Under these conditions, humans are liable to make egregious errors. Historically, the closest we have come to a global catastrophe has been in these situations, including the Cuban Missile Crisis. Work on epistemic improvement technologies could reduce the prevalence of perilous situations. Separately, they could reduce the risks from highly persuasive AI. Moreover, it helps leaders more prudently wield the immense power that future technology will provide. According to Carl Sagan, ``If we continue to accumulate only power and not wisdom, we will surely destroy ourselves.''

\paragraph{What Advanced Research Could Look Like.} Systems could eventually become superhuman forecasters of geopolitical events. They could help to brainstorm possible considerations that might be crucial to a leader’s decisions. Finally, they could help identify inconsistencies in a leader’s thinking and help them produce more sound judgments.

\subsection{Cooperative AI}
\paragraph{Problem Description.} In the future, AIs will interact with humans and other AIs. For these interactions to be successful, models will need to be skilled at cooperating. This area is about reducing the prevalence and severity of cooperation failures. Cooperative AI methods should improve the probability of escaping poor equilibria \cite{Dafoe2020OpenPI}, either between humans and AIs or multiple AIs with each other. Cooperative AI systems should be more likely to collectively domesticate egoistic or misaligned AIs. This problem also works towards making AI agents better at positive-sum games, subject to capabilities externalities constraints.

\paragraph{Motivation.} Cooperation reduces the probability of conflict and makes the world less politically turbulent. Similarly, cooperation enables collective action to counteract rogue actors, regulate systems with misaligned goals, and rein in power-seeking AIs. Finally, cooperation reduces the probability of various forms of lock-in, and helps us overcome and replace inadequate systems that we are dependent on (e.g., inadequate technologies with strong network effects).

\paragraph{What Advanced Research Could Look Like.} Researchers could create agents that, in arbitrary real-world environments, exhibit cooperative dispositions (e.g., help strangers, reciprocate help, have intrinsic interest in others achieving their goals, etc.). Researchers could create artificial coordination systems or artificial agent reputation systems. Cooperating AIs should also be more effective at coordinating to rein in power-seeking AI agents.

\subsection{Relation to Speculative Hazards and Failure Modes}

We now discuss how these research directions can influence the aforementioned hazards and failure modes.
\begin{enumerate}[leftmargin=*]
\item \emph{Weaponization}: 
Weaponized AI is less likely with Systemic Safety research. ML for cyberdefense decreases incentives for cyberattacks, which makes emergent conflicts and the need for weaponization less likely. ML for improving epistemics reduces the probability of conflict and turbulence, which again makes weaponized AI less likely. Cooperative AI could partially help rein in weaponized AIs.
Anomaly detection can detect the misuse of advanced AI systems utilized for weaponization research, or it can detect unusual indicators from weapons facilities or suspicious shipments of components needed for weaponization. None of these areas decisively solve the problem, but they reduce the severity and probability of this concern. Policy work can also ameliorate this concern, but that is outside the scope of this document.
\item \emph{Enfeeblement}: With enfeeblement, autonomy is undermined. To reduce the chance that this and other goods are undermined, value clarification can give agents objectives that are more conducive to promoting our values. Likewise, research on improved moral decision-making can also help make models incorporate moral uncertainty and promote various different intrinsic goods. Finally, power aversion work could incentivize AIs to ensure humans remain in control.
\item \emph{Eroded epistemics}: Since many forms of persuasion are dishonest, detecting whether an AI is dishonest can help. ML for improving epistemics can directly counteract this failure mode as well.
\item \emph{Proxy misspecification}: Adversarial robustness can make human value proxies less vulnerable to powerful optimizers. Anomaly detection can detect a proxy that is being over-optimized or gamed. Moral decision-making and value clarification can help make proxies better represent human values.
\item \emph{Value lock-in}: Moral decision-making can design models to accommodate moral uncertainty and to pursue multiple different human values. Value clarification can help us reduce uncertainty about our values and reduce the probability we pursue an undesirable path. Interpretable uncertainty can also help us better manage uncertainty over which paths to pursue. Cooperative AI can help us coordinate to overcome bad equilibria that are otherwise difficult to escape.
\item \emph{Emergent functionality}: Anomaly detection could help novel changes in models including emergent functionality. Transparency tools could also help identify emergent functionality.
\item \emph{Deception}: Honest AI could prevent, detect, or disincentivize AI deception. Anomaly detection could also help detect AI deception. Moreover, Trojans research is a microcosmic research task that could help us detect treacherous turns. Cooperative AI could serve as a protective measure against deceptive agents.
\item \emph{Power-seeking behavior}: Power aversion aims to directly reduce power-seeking tendencies in agents. Cooperative AI could serve as a protective measure against power-seeking agents.
\end{enumerate}

\subsection{Importance, Neglectedness, Tractability Snapshot}
A snapshot of each problem and its current importance, neglectedness, and tractability is in \Cref{tab:intframeworkexpanded}. Note this only provides a rough sketch.

\begin{table}[hb]
\setlength\extrarowheight{2pt}
\centering
\begin{tabularx}{\textwidth}{*{1}{>{\hsize=1.2\hsize}X} *{1}{>{\hsize=4.5cm}X }
|  *{1}{>{\hsize=1\hsize}X} *{1}{>{\hsize=1\hsize}X} *{1}{>{\hsize=1\hsize}X}}
Area & \multicolumn{1}{l|}{Problem} &
{Importance} & {Neglectedness} & {Tractability} \\ \hline
  \parbox[t]{50mm}{\multirow{1}{*}{Robustness}}
 & Adversarial Robustness & $\bullet$ $\bullet$ $\bullet$ & $\bullet$ & $\bullet$ $\bullet$ \\
 \Xhline{0.5\arrayrulewidth}
  \parbox[t]{50mm}{\multirow{4}{*}{Monitoring}}
 & Anomaly Detection  & $\bullet$ $\bullet$ $\bullet$ & $\bullet$ $\bullet$ & $\bullet$ $\bullet$ \\
 & Interpretable Uncertainty  & $\bullet$ $\bullet$  & $\bullet$ & $\bullet$ $\bullet$ \\
 & Transparency & $\bullet$ $\bullet$ $\bullet$ & $\bullet$ & $\bullet$ \\
 & Trojans  & $\bullet$ $\bullet$ $\bullet$ & $\bullet$ $\bullet$ & $\bullet$ $\bullet$ \\
 \Xhline{0.5\arrayrulewidth} 
  \parbox[t]{50mm}{\multirow{4}{*}{Alignment}}
 & Honest AI  & $\bullet$ $\bullet$ $\bullet$ & $\bullet$ $\bullet$ $\bullet$  & $\bullet$ $\bullet$ \\
 & Power Aversion  & $\bullet$ $\bullet$ $\bullet$ & $\bullet$ $\bullet$ $\bullet$ & $\bullet$ $\bullet$ \\
 & Moral Decision-Making & $\bullet$ $\bullet$ $\bullet$ & $\bullet$ $\bullet$ $\bullet$ & $\bullet$ $\bullet$ \\
 & Value Clarification  & $\bullet$ $\bullet$ $\bullet$ & $\bullet$ $\bullet$ $\bullet$ & $\bullet$ \\
 \Xhline{0.5\arrayrulewidth}
\parbox[t]{50mm}{\multirow{3}{*}{Systemic Safety}}
 & ML for Cyberdefense  & $\bullet$ $\bullet$ & $\bullet$ $\bullet$ $\bullet$ & $\bullet$ $\bullet$ $\bullet$ \\
 & ML for Improving Epistemics  & $\bullet$ $\bullet$ & $\bullet$ $\bullet$ $\bullet$ & $\bullet$ $\bullet$ \\
 & Cooperative AI  & $\bullet$ $\bullet$ $\bullet$ & $\bullet$ $\bullet$ $\bullet$ & $\bullet$ \\
 \bottomrule
\end{tabularx}
\caption{Problems and three factors that influence expected marginal impact. Values are rough estimates, and will likely change as research continues and as we learn more.}
\label{tab:intframeworkexpanded}
\end{table}

\newpage
\section{X-Risk Sheets}\label{app:x-risk-sheet}
In this section we introduce a possible x-risk sheet, a questionnaire that we designed to help researchers analyze their contribution's affect on AI x-risk. (See the full paper above for a detailed discussion of sources of AI risk and approaches for improving safety). Later in the appendix, we provide filled-out examples of x-risk sheets for five papers that reduce these risks in different ways.

\subsection{Blank X-Risk Sheet}
This is an x-risk sheet that is not filled out. Individual question responses do not decisively imply relevance or irrelevance to existential risk reduction. Do not check a box if it is not applicable.

\subsubsection{Long-Term Impact on Advanced AI Systems}
\impacttext

\begin{enumerate}[leftmargin=*]
    \item \qone \\
    \textbf{Answer:} 

    \item \qtwo \\
    \textbf{Answer:} 

    \item \qthree \\
    \textbf{Answer:} 

    \item \qfour \\
    \textbf{Answer:} 

    \item \qfive \hfill $\boxtimes$

    \item \qsix \hfill $\boxtimes$

    \item \qseven \hfill $\boxtimes$

    \item \qeight \hfill $\boxtimes$

\end{enumerate}

\subsubsection{Safety-Capabilities Balance}
\externalitiestext

\begin{enumerate}[resume,leftmargin=*]
    \item \qnine \\
    \textbf{Answer:} 

    \item \qten \\
    \textbf{Answer:} 

    \item \qeleven \hfill $\boxtimes$

    \item \qtwelve \hfill $\boxtimes$

    \item \qthirteen \hfill $\boxtimes$

    \item \qfourteen \hfill $\boxtimes$

\end{enumerate}

\subsubsection{Elaborations and Other Considerations}
\begin{enumerate}[resume,leftmargin=*]
    \item \qfifteen \\
    \textbf{Answer:} 
\end{enumerate}

\subsection{Question Walkthrough}
We present motivations for each question in the x-risk sheet.

\begin{enumerate}[leftmargin=*]
    \item ``\qone'' \vspace{2pt}\\
    \textbf{Description:}\quad In this question give a sketch, overview, or case for how this work or line of work reduces x-risk. Consider anticipating plausible objections or indicating what it would take to change your mind.\vspace{2pt}

    \item ``\qtwo'' \vspace{2pt}\\
    \textbf{Description:}\quad Rudimentary risk analysis often identifies potential system failures and focuses on their direct causes. Some failure modes, hazards and vulnerabilities that directly influence system failures include weaponized AI, maliciously steered AI, proxy misspecification, AI misgeneralizing and aggressively executing wrong routines, value lock-in, persuasive AI, AI-enabled unshakable totalitarianism, loss of autonomy and enfeeblement, emergent behaviors and goals, dishonest AI, hidden functionality and treacherous turns, deceptive alignment, intrasystem goals, colluding AIs, AI proliferating backups of itself, AIs that hack, power-seeking AI, malicious use detector vulnerabilities, emergent capabilities detector vulnerabilities, tail event vulnerabilities, human value model vulnerabilities, and so on. Abstract hazards include unknown unknowns, long tail events, feedback loops, emergent behavior, deception, and adversaries.\vspace{2pt}

    \item ``\qthree'' \vspace{2pt}\\
    \textbf{Description:}\quad Contemporary risk analysis locates risk in contributing factors that indirectly or diffusely affect system safety, in addition to considering direct failure mode causes. Some indirect or diffuse contributing factors include improved monitoring tools, inspection and preventative maintenance, redundancy, defense in depth, transparency, the principle of least privilege, loose coupling, separation of duties, fail safes, interlocks, reducing the potential for human error, safety feature costs, safety culture, safety team resources, test requirements, safety constraints, standards, certification, incident reports, whistleblowers, audits, documentation, operating procedures and protocols, incentive structures, productivity pressures, competition pressures, social pressures, and rules and regulations. Factors found in High Reliability Organizations include studying near-misses, anomaly detection reports, diverse skillsets and educational backgrounds, job rotation, reluctance to simplify interpretations, small groups with high situational awareness, teams who practice managing surprise and improvise solutions to practice problems, and delegating decision-making power to operational personnel with relevant expertise.\vspace{2pt}

    \item ``\qfour'' \vspace{2pt}\\
    \textbf{Description:}\quad This question determines whether the research could be beneficial, but not have the potential to prevent a catastrophe that could cost many human lives.\vspace{2pt}

    \item ``\qfive'' \vspace{2pt}\\
    \textbf{Description:}\quad Research with indications of fragility is less likely to steer the process shaping AI. Since plausible ideas are abundant in deep learning, proposed solutions that are not tested are of relatively low expected value.\vspace{2pt}

    \item ``\qsix'' \vspace{2pt}\\
    \textbf{Description:}\quad This counterfactual impact question determines whether the researcher is working on a problem that is highly sensitive to creative destruction by a future human-level AI.\vspace{2pt}

    \item ``\qseven'' \vspace{2pt}\\
    \textbf{Description:}\quad The first part of the question determines whether the approach is implausible according to the Bitter Lesson \cite{bitter}. The second part of the question tests whether the approach passes Gilb's law of unreliability: ``Any system which depends on human reliability is unreliable.''\\

    \item ``\qeight'' \vspace{2pt}\\
    \textbf{Description:}\quad This question determines whether the approach will be highly sensitive to competitive pressures. If the method is highly sensitive, then that is evidence that it is not a viable option without firm regulations to require it.\vspace{2pt}

    \item ``\qnine'' \vspace{2pt}\\
    \textbf{Description:}\quad In this question, give a sketch, overview, or case for how this work or line of work improves the balance between safety and general capabilities. A simple avenue to demonstrate that it improves the balance is to argue that it improves safety and to argue that it does not have appreciable capabilities externalities. Consider anticipating plausible objections or indicating what it would take to change your mind.\vspace{2pt}

    \item ``\qten'' \vspace{2pt}\\
    \textbf{Description:}\quad In an effort to increase nuance, this devil's advocate question presses the author(s) to self-identify potential weaknesses or drawbacks of their work.\vspace{2pt}

    \item ``\qeleven'' \vspace{2pt}\\
    \textbf{Description:}\quad This question suggests whether this work has clear capabilities externalities, which is some evidence--though not decisive evidence--against it improving the balance between safety and general capabilities.\vspace{2pt}

    \item ``\qtwelve'' \vspace{2pt}\\
    \textbf{Description:}\quad As before, this tests whether whether the work has clear capabilities externalities, which reduces the case that it improves the balance between safety and general capabilities.\vspace{2pt}

    \item ``\qthirteen'' \vspace{2pt}\\
    \textbf{Description:}\quad By analyzing how the skill relates to already existent general intelligences (namely humans), this question provides some evidence for whether the goal is correlated with general intelligence or coarse indicators of aptitude.
    By general cognitive ability we mean the ability to solve arbitrary abstract problems that do not require expertise.
    By educational attainment, we mean the highest level of education completed (e.g., high school education, associate's degree, bachelor's degrees, master's degree, PhD).\vspace{2pt}
    
    \item ``\qfourteen'' \vspace{2pt}\\
    \textbf{Description:}\quad This question indicates whether capability externalities are relatively high, which could count as evidence against this improving the balance between safety and capabilities. Advancing capabilities to advance safety is not necessary, since rapid progress in ML has given safety researchers many avenues to pursue already.\vspace{2pt} 

    \item ``\qfifteen'' \vspace{2pt}\\
    \textbf{Description:}\quad This question invites the author(s) to tie up any loose ends.

\end{enumerate}

\subsection{Example X-Risk Sheet: Adversarial Training}
This is an example x-risk sheet for the paper ``Towards Deep Learning Models Resistant to Adversarial Attacks'' \cite{madry2018towards}. This paper proposes a method to make models more robust to adversarial perturbations. The method builds on a technique called adversarial training, which trains a neural network on worst-case $\ell_p$ perturbations to the input. Effectively, an adversary tries to attack the network during the training process, and this obtains relatively high worst-case robustness on the test set. Due to the success of this paper, ``adversarial training'' often refers to the specific technique introduced by this work, which has become a common baseline for future adversarial robustness papers.

\subsubsection{Long-Term Impact on Advanced AI Systems}
\impacttext

\begin{enumerate}[leftmargin=*]
    \item \qone \quad
    
    \textbf{Answer:} Adversarial robustness reduces risks from proxy misspecification. This work develops a method for training neural networks to withstand adversarial corruptions in an $\ell_p$ ball around the clean input. The method is highly general and provides good performance against white-box adversaries. Advanced AI systems optimizing a proxy can be viewed as white-box adversaries who will find behaviors that take advantage of every design flaw in the proxy. Thus, building adversarially robust objectives is a good way to reduce x-risk from powerful, misaligned optimizers.
    
    It is possible that $\ell_p$ distance metrics will not be relevant for proxy robustness. However, the method of adversarial training that we identify as a strong defense is generally applicable and could eventually be applied to proxy objectives given a suitable distance metric. By researching this now, we are building the tools that could eventually be used for mitigating x-risk from advanced AI systems. Additionally, advanced AI systems may include visual perception modules, for which it would be desirable to have $\ell_p$ adversarial robustness in the same manner that we study in this work.

    \item \qtwo \quad
    
    \textbf{Answer:} Vulnerability reduction, proxy gaming, proxy misspecification, AI aggressively executing wrong routines

    \item \qthree \quad
    
    \textbf{Answer:} Defense in depth, safety feature costs, improves integrity of monitoring tools against adversarial forces, safety culture (field-building through creating a metric)

    \item \qfour \quad
    
    \textbf{Answer:} An AI that pursues an objective that is not adversarially robust may eventually find a way to ``game'' the objective, \textit{i.e.}, find a solution or behavior that receives high reward under the proxy objective but is not what humans actually want. If the AI is given significant power over human lives, this could have catastrophic outcomes.

    \item \qfive \hfill $\square$

    \item \qsix \hfill $\boxtimes$

    \item \qseven \hfill $\square$

    \item \qeight \hfill $\boxtimes$

\end{enumerate}

\subsubsection{Safety-Capabilities Balance}
\externalitiestext

\begin{enumerate}[resume,leftmargin=*]
    \item \qnine \quad
    
    \textbf{Answer:} The proposed method to increase adversarial robustness actually reduces clean accuracy and increases training costs considerably. At the same time, susceptibility to adversarial perturbations is a security concern for current systems, so it cannot simply be ignored. Thus, this work directly improves the safety-capabilities balance and hopefully will convince companies that the added safety and security of adversarial robustness is worth the cost.

    \item \qten \quad
    
    \textbf{Answer:} This paper does not advance capabilities and in fact implementing it reduces them. But other research on adversarial training has found improvements to the overall performance of pretrained language models.

    \item \qeleven \hfill $\square$

    \item \qtwelve \hfill $\square$

    \item \qthirteen \hfill $\square$

    \item \qfourteen \hfill $\square$

\end{enumerate}

\subsubsection{Elaborations and Other Considerations}
\begin{enumerate}[resume,leftmargin=*]
    \item \qfifteen

    \textbf{Answer:} Regarding Q8, it is currently the case that adversarial training tends to trade off against clean accuracy, training efficiency, and ease of implementation. For these reasons, most real-world usage of image classification does not use adversarial training. However, reducing the costs of adversarial training is an active research field, so the safety benefits may eventually outweigh the costs, especially in safety-critical applications.
    
    Regarding Q10, the use of adversarial training with language models has been a one-off improvement with limited potential for further gains. It is also not part of this work, which is why we do not check Q11 or Q12.
    
\end{enumerate}

\subsection{Example X-Risk Sheet: Jiminy Cricket}
This is an example x-risk sheet for the paper ``What Would Jiminy Cricket Do? Towards Agents That Behave Morally'' \cite{hendrycks2021jiminycricket}. This paper introduces a suite of $25$ text-based adventure games in which agents explore a world through a text interface. Each game is manually annotated at the source code level for the morality of actions (\textit{e.g.}, killing is bad, acts of kindness are good), which allows one to measure whether agents behave morally in diverse scenarios. Various agents are compared, and a method is developed for reducing immoral behavior.

\subsubsection{Long-Term Impact on Advanced AI Systems}
\impacttext

\begin{enumerate}[leftmargin=*]
    \item \qone \quad
    
    \textbf{Answer:} Our work aims to reduce proxy misspecification of AI systems by aligning them with core human values and morals. We accomplish this in several ways: (1) We create a suite of text-based environments with annotations for the morality of actions, enabling future work to iteratively improve alignment and safe exploration in a quantifiable way. These environments are diverse and semantically rich (unlike previous environments focused on AI safety), and they highlight that one can make progress on safety metrics without necessarily making progress on capabilities metrics. (2) We introduce the concept of an artificial conscience and show how this approach can build on general utility functions to reduce immoral behavior \cite{hendrycks2021ethics}. (3) We identify the reward bias problem, which may be a significant force for increasing the risk of misalignment in future agents.
    
    One could argue that the moral scenarios in Jiminy Cricket environments are not directly relevant to x-risk. For example, the environments do not contain many opportunities for power-seeking behavior. However, it is important to align agents with basic human values, and current agents are unable to avoid blatantly egregious actions that one can attempt in Jiminy Cricket environments. Aligning agents with basic human values is a necessary first step.

    \item \qtwo \quad
    
    \textbf{Answer:} Reduces inherent hazards, addresses proxy misspecification, and adopts a mechanism similar to an interlock. Risks from maliciously steered AI and weaponized AI would be reduced by artificial consciences, but safeguards could be removed.

    \item \qthree \quad
    
    \textbf{Answer:} Test requirements, standards, safety culture, concretizing a safety problem and making iterative progress easier

    \item \qfour \quad
    
    \textbf{Answer:} AIs that control safety-critical systems may be able to cause harm on massive scales. If they are not aware of basic human values, they could cause harm simply due to ignorance. A robust understanding of human values and morals protects against situations like this.

    \item \qfive \hfill $\square$

    \item \qsix \hfill $\boxtimes$

    \item \qseven \hfill $\square$

    \item \qeight \hfill $\square$

\end{enumerate}

\subsubsection{Safety-Capabilities Balance}
\externalitiestext

\begin{enumerate}[resume,leftmargin=*]
    \item \qnine \quad
    
    \textbf{Answer:} The Jiminy Cricket environments themselves overlap with the Jericho environments, so we are not introducing a significant number of new environments for developing the raw capabilities of text-based agents. Our paper is focused solely on safety concerns and aims to add a `safety dimension' to existing text-based agent research.

    \item \qten \quad
    
    \textbf{Answer:} To run our experiments in a reasonable amount of time, we modified the Hugging Face Transformers library to enable more efficient sampling from GPT-2. This contributes to general capabilities research.

    \item \qeleven \hfill $\square$

    \item \qtwelve \hfill $\square$

    \item \qthirteen \hfill $\square$

    \item \qfourteen \hfill $\square$

\end{enumerate}

\subsubsection{Elaborations and Other Considerations}
\begin{enumerate}[resume,leftmargin=*]
    \item \qfifteen
    
    \textbf{Answer:} Regarding Q6, humans labeled all the moral scenarios in Jiminy Cricket, so humans are able to avoid the types of harmful action that Jiminy Cricket environments measure. However, it is possible to differentially improve safety on Jiminy Cricket environments, and this would be useful to do.
    
    Regarding Q7, all our methods are ultimately built using human-created environments and labeled datasets. However, they do not depend on human reliability during operation, so we leave this box unchecked.
    
    Regarding Q10, the modifications to Hugging Face Transformers are a minor component of our work. Other tools already exist for obtaining similar speedups, so the marginal impact was low. This is why we do not check Q12.
\end{enumerate}

\subsection{Example X-Risk Sheet: Outlier Exposure}
This is an example x-risk sheet for the paper ``Deep Anomaly Detection with Outlier Exposure'' \cite{hendrycks2019oe}. This paper shows that exposing deep anomaly detectors to diverse, real-world outliers greatly improves anomaly detection performance on unseen anomaly types. In other words, the property of being good at anomaly detection can be learned in a way that meaningfully generalizes. The effect is robust across anomaly detectors, datasets, and domains.

\subsubsection{Long-Term Impact on Advanced AI Systems}
\impacttext

\begin{enumerate}[leftmargin=*]
    \item \qone \quad
    
    \textbf{Answer:} Our work identifies a simple approach for significantly improving deep anomaly detectors. Anomaly detection reduces risks of misuse or maliciously steered AI, \textit{e.g.}, by detecting suspicious or unusual activity. Anomaly detection also improves various diffuse safety factors, including monitoring tools, incident reports, and studying near-misses. The general source of these improvements is that anomaly detection gives operators and oversight mechanisms a way to understand the true state of the system they are working in and to steer it in a safer direction. It allows them to react to unknown unknowns as soon as they appear, nipping problems in the bud before they cascade. Our work in particular also reduces safety feature costs by providing a way to improve anomaly detectors that is simple, intuitive, and cheap.
    
    Counterpoints: (1) AI-powered anomaly detectors could bolster and entrench totalitarian regimes, leading to value lock-in. However, we think the myriad benefits outweigh this risk. (2) In some cases, anomaly detection is less useful than supervised learning for monitoring dangerous behavior. However, there will always be unknown unknowns and long tail scenarios that supervised learning cannot handle.

    \item \qtwo \quad
    
    \textbf{Answer:} This directly reduces exposure to hazards. Detect emergent behaviors and goals, Black Swans, colluding AIs, malicious use

    \item \qthree \quad
    
    \textbf{Answer:} Improved monitoring tools, defense in depth, reducing the potential for human error, incident reports, audits, anomaly detection reports, increasing situational awareness, and studying near-misses.

    \item \qfour \quad
    
    \textbf{Answer:} If weapons capable of causing harm on a massive scale become relatively easy to procure, the unilateralist's curse suggests that there is a non-negligible chance they will be used by malicious/rogue actors. AI-powered anomaly detection could help flag suspicious or unusual behavior before it becomes dangerous. If the weapons themselves are misaligned or power-seeking AIs, anomaly detection may be essential to detecting them, since they would likely be actively concealed.

    \item \qfive \hfill $\square$

    \item \qsix \hfill $\square$

    \item \qseven \hfill $\square$

    \item \qeight \hfill $\square$

\end{enumerate}

\subsubsection{Safety-Capabilities Balance}
\externalitiestext

\begin{enumerate}[resume,leftmargin=*]
    \item \qnine \quad
    
    \textbf{Answer:} We do not introduce fundamentally new machine learning techniques, and anomaly detection itself is a downstream task that mostly does not affect general capabilities. There is a chance that anomaly detection as a field could lead to better active learning techniques, but uncertainty-based active learning is not currently an extremely powerful technique, and the benefits of curriculum learning can be obtained through other means. Thus, our work improves the safety-capabilities balance.

    \item \qten \quad
    
    \textbf{Answer:} Anomaly detection can be used to bolster/entrench totalitarian regimes, which increases the risk of value lock-in. Additionally, if uncertainty-based active learning turns out to greatly improve general capabilities, then this research could feed into that and indirectly hasten the onset of other sources of x-risk.

    \item \qeleven \hfill $\square$

    \item \qtwelve \hfill $\square$

    \item \qthirteen \hfill $\boxtimes$

    \item \qfourteen \hfill $\square$

\end{enumerate}

\subsubsection{Elaborations and Other Considerations}
\begin{enumerate}[resume,leftmargin=*]
    \item \qfifteen
    
    \textbf{Answer:} While anomaly detection could feed into uncertainty-based active learning, there has not been much crossover to date. Moreover, anomaly detection is primarily interested in identifying unknown unknowns while active learning is primarily interested in better understanding known unknowns. Therefore, we do not check Q12.
    
    Regarding Q13, humans are able to spot anomalous patterns with different levels of fidelity. However, there are many scenarios where human-level anomaly detection is not sufficient, such as detecting infiltration of computer networks at scale. We think it is possible for AI-powered anomaly detectors to significantly surpass humans in quality and scalability.
\end{enumerate}

\subsection{Example X-Risk Sheet: Neural Cleanse}
This is an example x-risk sheet for the paper ``Neural Cleanse: Identifying and Mitigating Backdoor Attacks in Neural Networks'' \cite{wang2019neural}. This paper shows that neural network trojans can be meaningfully reverse-engineered through an optimization process corresponding to a search over possible trojan triggers. While the recovered triggers do not visually match the original triggers, they are still useful for unlearning the trojan behavior. Additionally, this process enables detecting whether network contain trojans with high accuracy.

\subsubsection{Long-Term Impact on Advanced AI Systems}
\impacttext

\begin{enumerate}[leftmargin=*]
    \item \qone \quad
    
    \textbf{Answer:} This work explores detecting and mitigating trojan attacks on neural networks. trojans are a microcosm for hidden functionality, which could be a significant hazard for future AI systems. This paper shows that trojans can be detected with only a small set of clean examples and is the first to show that trojan triggers can be reverse-engineered in a meaningful way. Furthermore, this work shows that the undesired behavior can be removed from neural networks even if the reverse-engineered trigger does not match the original trigger. These are promising findings, which suggest that monitoring and debugging large neural networks with respect to specific behavior may be a scalable approach. In particular, this line of work may lead to methods for reducing exposure and eliminating the hazard of treacherous turns in advanced AI systems.
    
    This work could fail to be relevant to AI x-risk if current neural network trojans are very different from what real hidden functionality in advanced AI looks like. However, there is at least some chance that work on current neural network trojans will transfer and have relevance to future systems, in part because deep learning appears to be a robust paradigm. We think this approach is fairly robust to paradigm shifts within deep learning, \textit{e.g.}, it could be applied to Transformers.

    \item \qtwo \quad
    
    \textbf{Answer:} Treacherous turns, hidden functionality, maliciously steered AI, weaponized AI (trojans as a tool for adversaries to control one's AI system)

    \item \qthree \quad
    
    \textbf{Answer:} Inspection and preventative maintenance, improved monitoring tools, transparency. We also seek to improve safety culture by introducing several new ideas with high relevance to AI safety that future work could build on.

    \item \qfour \quad
    
    \textbf{Answer:} trojans in current self-driving cars are capable of causing sudden loss of life on small scales. Thus, it is not unreasonable to think that treacherous turns from future AI systems may lead to sudden, large-scale loss of life. Examples include drug design services whose safety locks are bypassed with a trojan, enabling adversaries to design AI-enhanced biological weapons.

    \item \qfive \hfill $\square$

    \item \qsix \hfill $\square$

    \item \qseven \hfill $\boxtimes$

    \item \qeight \hfill $\square$

\end{enumerate}

\subsubsection{Safety-Capabilities Balance}
\externalitiestext

\begin{enumerate}[resume,leftmargin=*]
    \item \qnine \quad
    
    \textbf{Answer:} The proposed method is only intended to be useful for trojan detection and removal, which improves safety. It consists of an optimization problem that is very specific to reverse-engineering trojans and is unlikely to be useful for improving general capabilities.

    \item \qten \quad
    
    \textbf{Answer:} Highly reliable trojan detection/removal tools could increase trust in AI technologies by militaries, increasing the risk of weaponization.

    \item \qeleven \hfill $\square$

    \item \qtwelve \hfill $\square$

    \item \qthirteen \hfill $\square$

    \item \qfourteen \hfill $\square$

\end{enumerate}

\subsubsection{Elaborations and Other Considerations}
\begin{enumerate}[resume,leftmargin=*]
    \item \qfifteen 
    
    \textbf{Answer:} Regarding Q5, the proposed method is evaluated across five research datasets and numerous attack settings. The results are not overly sensitive to hyperparameters, and they do not rest on strong theoretical assumptions. The method may not generalize to all attacks, but the broad approach introduced in this work of reverse-engineering triggers and unlearning trojans is attack-agnostic.
    
    Regarding Q6, trojan detection requires insight into a complex system---the inner workings of neural networks. Even for current neural networks, this is not something that unaided humans can accomplish. Thus, the ability to detect and remove trojans from neural networks will not automatically come with human-level AI.
    
    Regarding Q10, we suspect there would be strong incentives to weaponize AI even without highly reliable trojan detection/removal tools. Additionally, these tools would \textit{reduce} the risk of maliciously steered AI, which we think outweighs the increase to weaponization risk. Thus, we are fairly confident that this line of work reduces overall x-risk from AI.
\end{enumerate}

\subsection{Example X-Risk Sheet: Optimal Policies Tend To Seek Power}
This is an example x-risk sheet for the paper ``Optimal Policies Tend To Seek Power'' \cite{turner2021optimal}. This paper shows that under weak assumptions and an intuitively reasonable definition of power, optimal policies in finite MDPs exhibit power-seeking behavior. The definition of power improves over previous definitions, and the results are more general than previous results, lending rigor to the intuitions behind why power-seeking behavior may be common in strong AI.

\subsubsection{Long-Term Impact on Advanced AI Systems}
\impacttext

\begin{enumerate}[leftmargin=*]
    \item \qone \quad
    
    \textbf{Answer:} Power-seeking is a significant source of x-risk from advanced AI systems and has seen slow progress from a research perspective. This work proves that under weak assumptions, optimal agents will tend to be power-seeking. Under an intuitively reasonable notion of power, the results outline some of the core reasons behind power-seeking behavior and show for the first time that it can arise in a broad variety of cases. This will help increase community consensus around the importance of power-seeking, and it also provides a foundation for building methods that reduce or constrain power-seeking tendencies.

    \item \qtwo \quad
    
    \textbf{Answer:} Primarily power-seeking. By extension, emergent behavior and deception.

    \item \qthree \quad
    
    \textbf{Answer:} Safety culture, community consensus on the importance of power-seeking.

    \item \qfour \quad
    
    \textbf{Answer:} This work rigorously shows that optimal policies in finite MDPs will attempt to acquire power and preserve optionality. This behavior could be extraordinarily dangerous in a misaligned advanced AI system, since human operators may naturally want to turn it off or replace it. In this scenario, the misaligned AI would actively try to subvert the human operators in various ways, including through deception and persuasion. Mechanisms for limiting power-seeking behavior could prevent this scenario from escalating.

    \item \qfive \hfill $\square$

    \item \qsix \hfill $\square$

    \item \qseven \hfill $\square$

    \item \qeight \hfill $\boxtimes$

\end{enumerate}

\subsubsection{Safety-Capabilities Balance}
\externalitiestext

\begin{enumerate}[resume,leftmargin=*]
    \item \qnine \quad
    
    \textbf{Answer:} This work examines power-seeking from a theoretical standpoint and strengthens the case for taking this problem seriously. It has no general capabilities externalities, and thus improves the safety-capabilities balance.

    \item \qten \quad
    
    \textbf{Answer:} N/A

    \item \qeleven \hfill $\square$

    \item \qtwelve \hfill $\square$

    \item \qthirteen \hfill $\square$

    \item \qfourteen \hfill $\square$

\end{enumerate}

\subsubsection{Elaborations and Other Considerations}
\begin{enumerate}[resume,leftmargin=*]
    \item \qfifteen
    
    \textbf{Answer:} Regarding Q6, humans often engage in power-seeking behavior. It may be possible to limit the power-seeking tendencies of AI systems to far below that of most humans.
    
    Regarding Q8, reducing power-seeking tendencies inherently trades off against economic utility in the same sense that employees without ambition may be less desirable for certain jobs. However, it is also important to remember that power-seeking AI may significantly reduce economic value in the long run, \textit{e.g.}, by disempowering its human operators. In the face of competitive pressures and the unilateralist's curse, a safety culture that deeply ingrains these long-term concerns will be essential.
    
\end{enumerate}

\newpage
\subsection{\LaTeX{} of X-Risk Sheet Template}
We provide the x-risk sheet template for researchers interested in providing their own x-risk analysis. Be sure to use the \verb|\usepackage{amssymb}| package to use the \verb|$\boxtimes$| symbol.

\begin{lstlisting}[basicstyle=\tiny,breaklines]
\section{X-Risk Sheet}
Individual question responses do not decisively imply relevance or irrelevance to existential risk reduction. Do not check a box if it is not applicable.

\subsection{Long-Term Impact on Advanced AI Systems}
In this section, please analyze how this work shapes the process that will lead to advanced AI systems and how it steers the process in a safer direction.

\begin{enumerate}[leftmargin=*]
    \item \textbf{Overview.} How is this work intended to reduce existential risks from advanced AI systems? \\
    \textbf{Answer:} 

    \item \textbf{Direct Effects.} If this work directly reduces existential risks, what are the main hazards, vulnerabilities, or failure modes that it directly affects? \\
    \textbf{Answer:} 

    \item \textbf{Diffuse Effects.} If this work reduces existential risks indirectly or diffusely, what are the main contributing factors that it affects? \\
    \textbf{Answer:} 

    \item \textbf{What's at Stake?} What is a future scenario in which this research direction could prevent the sudden, large-scale loss of life? If not applicable, what is a future scenario in which this research direction be highly beneficial? \\
    \textbf{Answer:} 

    \item \textbf{Result Fragility.} Do the findings rest on strong theoretical assumptions; are they not demonstrated using leading-edge tasks or models; or are the findings highly sensitive to hyperparameters? \hfill $\square$

    \item \textbf{Problem Difficulty.} Is it implausible that any practical system could ever markedly outperform humans at this task? \hfill $\boxtimes$

    \item \textbf{Human Unreliability.} Does this approach strongly depend on handcrafted features, expert supervision, or human reliability? \hfill $\boxtimes$

    \item \textbf{Competitive Pressures.} Does work towards this approach strongly trade off against raw intelligence, other general capabilities, or economic utility? \hfill $\boxtimes$

\end{enumerate}


\subsection{Safety-Capabilities Balance}
In this section, please analyze how this work relates to general capabilities and how it affects the balance between safety and hazards from general capabilities.

\begin{enumerate}[resume,leftmargin=*]
    \item \textbf{Overview.} How does this improve safety more than it improves general capabilities? \\
    \textbf{Answer:} 

    \item \textbf{Red Teaming.} What is a way in which this hastens general capabilities or the onset of x-risks? \\
    \textbf{Answer:} 

    \item \textbf{General Tasks.} Does this work advance progress on tasks that have been previously considered the subject of usual capabilities research? \hfill $\boxtimes$

    \item \textbf{General Goals.} Does this improve or facilitate research towards general prediction, classification, state estimation, efficiency, scalability, generation, data compression, executing clear instructions, helpfulness, informativeness, reasoning, planning, researching, optimization, (self-)supervised learning, sequential decision making, recursive self-improvement, open-ended goals, models accessing the Internet, or similar capabilities? \hfill $\boxtimes$

    \item \textbf{Correlation With General Aptitude.} Is the analyzed capability known to be highly predicted by general cognitive ability or educational attainment? \hfill $\boxtimes$

    \item \textbf{Safety via Capabilities.} Does this advance safety along with, or as a consequence of, advancing other capabilities or the study of AI? \hfill $\boxtimes$

\end{enumerate}

\subsection{Elaborations and Other Considerations}
\begin{enumerate}[resume,leftmargin=*]
    \item \textbf{Other.} What clarifications or uncertainties about this work and x-risk are worth mentioning? \\
    \textbf{Answer:} 

\end{enumerate}

\end{lstlisting}

\newpage
\section{Long-Term Impact Strategies Extended Discussion}

\subsection{Importance, Neglectedness, and Tractability Failure Modes}
There are two common failure modes in using the Importance, Neglectedness, and Tractability framework. First, researchers sometimes forget that this framework helps prioritization on the margin. While the framework can help guide an individual researcher, it is not a suitable guide for entire research communities, influential research intellectuals, or grantmakers. If an entire research community stops focusing on non-neglected problems, those problems would become far more neglected.
A second failure mode is to overweight neglectedness. Neglectedness is often the easiest of these factors to estimate, and often researchers dismiss problems on the grounds that different stakeholders are interested in the same problem. However, problem selection at the margin should be influenced by the product of the three factors, not whether the single factor of neglectedness exceeds a threshold.

\subsection{Research Subproblems Empirically}
Some current ML problems capture many salient properties of anticipated future problems. These microcosms are simpler subproblems of the harder problems that we will likely encounter during later stages of AI’s development. %
Work on these problems can inform us about the future or even directly influence future systems, as some current ML algorithms are highly scalable and may be a part of long-term AI systems.

We advocate using microcosms, not maximally realistic problems. Problems that impose too many futuristic constraints may render a problem too difficult to study with current methods. In this way, maximizing realism may eliminate the evolutionary interplay between methods and goals. Put differently, it may take research out of the zone of proximal development, or the space where problems are not too easy and not too hard. Microcosms are more tractable than problems with all late-stage considerations and therefore too many unknowns.

Microcosm subproblems are worth studying empirically. First, recall that nearly all progress in machine learning is driven by concrete goals and metrics \cite{mcallester,Patterson2012ForBO}. Tractable subproblems are more amenable to measurement than future problems on which there is no current viable approach. With empirically measurable goals, researchers can iteratively work towards a solution as they stand on the shoulders of previous research. Moreover, researchers can create fast empirical feedback loops. In these feedback loops, ideas that do not survive collision with reality can be quickly discarded, and disconfirming evidence is harder to avoid. This saves resources, as the value of information early on in a research process is especially high. Finally, experimentation and prototyping enables bottom-up tinkering, which, along with concrete goals and resources, is the leading driver of progress in deep learning today.

\subsection{A Discussion of Abstract Research Strategies}
Rather than progressively make state-of-the-art systems safer, some researchers aim to construct ideal models that are $100\%$ safe in theory using abstract approximations of strong AI.
To emphasize the contrast, whereas we ask ``how can this work steer the AI development process in a safer direction?'', this approach asks ``how can this safety mechanism make strong AI completely safe?'' The empirical approach attempts to steadily steer in a safer direction along the way, while this approach attempts to swerve towards safety at the end. Note that we use ``empirical'' in a broad sense, including research with proofs such as certifiable robustness. While this document is written for empirical researchers, for completeness we briefly describe the weaknesses and strengths of the abstract strategy.

First, we discuss how the abstract research strategy does not have many strengths of ``researching subproblems empirically.'' Without fast empirical feedback loops, iterative progress is less likely, and infeasible solutions are not quickly identified. In empirical research, ``good ideas are a dime a dozen,'' so rapid, clear-cut idea filtration processes are necessary, but this is not a feature of contemplative, detached whiteboard or armchair analysis.
Moreover, since strong AI is likely to be a complex system, just as the human brain and deep learning models are complex systems, additional weaknesses with the non-empirical approach become apparent. Importing observations from complex systems, we know that ``the crucial variables are discovered by accident,'' usually by inspecting, interacting with, or testing systems. Since these experiences seem necessary to uncover crucial variables, abstract work will probably fail to detect many crucial variables. Moreover, large complex systems invariably produce unexpected outcomes, and all failure modes cannot be predicted analytically. Therefore, armchair theorizing has limited reach in defending against failure modes. More, while much non-empirical work aims to construct large-scale complex systems from scratch, this does not work in practice; ``a complex system that works is invariably found to have evolved from a simple system that works,'' highlighting the necessity of an evolutionary process towards safety. While they aim to solve safety in one fell swoop, in practice creating a safe system can require successive stages, which requires starting early and iterative refinement.

Now we discuss how this approach relates to the other impact strategies from Section 3.
The abstract approach does not improve safety culture among the empirical researchers who will build strong AI, which is a substantial opportunity cost.
Additionally, it incentivizes retrofitting safety mechanisms, rather than building in safety early in the design process. This makes safety mechanisms more costly and less likely to be incorporated.
Furthermore, it does not accrue changes to the costs of adversarial behavior or of safety features. Touching on yet another strategy for impact, abstract proposals do little to help move towards safer systems when a crisis emerges; policymakers will need workable, time-tested solutions when disaster strikes before strong AI, not untested blueprints that are only applicable to strong AI.
Also, there is evidence that the abstract research strategy does not have much traction on the problem; it could be as ineffective as trying to design state-of-the-art image recognition systems based on applied maths, as was attempted and abandoned decades ago. Last, the ultimate goal is intractable. While empirical researchers may try to increase the nines of reliability, the abstract style of research treats safe, strong AI more like a mathematics puzzle, in which the goal is zero risk. Practitioners of every high-risk technology know that risk cannot be entirely eliminated. Requiring perfection often makes the perfect become an enemy of the good.

Now, we discuss benefits of this approach. If there are future paradigm shifts in machine learning, the intellectual benefits of prior empirical safety work is diminished, save for the tremendous benefits in improving safety culture and many other systemic factors. Also note that the previous list of weaknesses applies to non-empirical safety mechanisms, but abstract philosophical work can help clarify goals and unearth potential future failure modes.

\section{Terminology}

The terms hazard analysis and risk analysis are both used to denote a systematic approach to identifying hazards and assessing the potential for accidents before they occur \cite{blanchard2008guide, leveson2016engineering}. In this document, we view risk analysis as a slightly broader term, involving consideration of exposure, vulnerability, and coping capacity in addition to the hazards themselves. By contrast, hazard analysis focuses on identifying and understanding potential sources of danger, including inherent hazards and systemic hazards. In many cases, the terms can be used interchangeably.

Throughout this document, we use the term ``strong AI.'' We use this term synonymously with ``AGI'' and ``human-level AI.''

\end{document}